%% file: main.tex
\documentclass[conference]{IEEEtran}
\IEEEoverridecommandlockouts

\usepackage[T1]{fontenc}
\usepackage{cite}
\usepackage{amsmath,amssymb,amsfonts}
\usepackage{algorithmic}
\usepackage{fontawesome}
\usepackage{graphicx}
\usepackage[table,dvipsnames]{xcolor}
\usepackage{textcomp}
\DeclareTextCommandDefault{\textquotedbl}{\char34\relax}
\DeclareTextCommandDefault{\textquotesingle}{\char39\relax}
\usepackage{tikz}
\usepackage{hyperref}
\usepackage{tcolorbox}
\usepackage{tabularx}
\usepackage{makecell}
\usepackage{multirow}
\usepackage{hhline}
\usepackage{relsize}
\usepackage{booktabs}
\usepackage{listings}
\usepackage{xspace}
\usepackage{paralist}
\usepackage{stfloats}
\usepackage[normalem]{ulem}

\usetikzlibrary{calc, fit, backgrounds}
\def\BibTeX{{\rm B\kern-.05em{\sc i\kern-.025em b}\kern-.08em
    T\kern-.1667em\lower.7ex\hbox{E}\kern-.125emX}}

\definecolor{veryLightGray}{RGB}{248,248,248}
\definecolor{codegray}{RGB}{245,245,245}

\newcommand{\etal}{\textit{et al.}\xspace}
\newcommand{\ie}{\textit{i.e.,}\xspace}
\newcommand{\eg}{\textit{e.g.,}\xspace}

\newcommand{\secref}[1]{Section~\ref{#1}\xspace}

\newcommand{\figref}[1]{Figure~\ref{#1}\xspace}

\newcommand{\tabref}[1]{Table~\ref{#1}\xspace}
\newcommand{\pharo}{\textit{Pharo}\xspace}
\newcommand{\compl}{\textsc{Complishon}\xspace}
\newcommand{\SFT}[0]{\textit{SFT}\xspace}
\newcommand{\PT}[0]{\textit{PT}\xspace}

\renewcommand{\tabularxcolumn}[1]{m{#1}}
\newcolumntype{C}{>{\centering\arraybackslash}X}

\newenvironment{findbox}{
  \par%
  \begingroup
  \setlength{\fboxsep}{6pt}%
  \hspace{-0.4cm}%
  \setbox0=\vbox\bgroup\noindent
  \hsize=0.95\linewidth
  \begin{minipage}{0.95\linewidth}\normalsize
}{%
  \end{minipage}\egroup
  \begingroup
    \fboxrule=0.6pt
    \fcolorbox{black!80}{gray!10}{\box0}%
  \endgroup
  \endgroup\par\noindent
  \normalcolor\ignorespacesafterend
}

\begin{document}

\title{Teaching LLMs a Low-Resource Language:\\Enhancing Code Completion in Pharo}

\author{
\begingroup
\small
\setlength{\tabcolsep}{3pt}
\renewcommand{\arraystretch}{1.08}

\begin{tabular}{>{\centering\arraybackslash}p{0.31\textwidth}
                >{\centering\arraybackslash}p{0.31\textwidth}
                >{\centering\arraybackslash}p{0.31\textwidth}}

Kilian Kier$^\ast$ &
Alessandro Giagnorio$^\ast$ &
Omar AbedelKader$^\ast$ \\
kilian.kier@student.tugraz.at &
alessandro.giagnorio@usi.ch &
omar.abedelkader@inria.fr \\
\textit{Graz University of Technology} &
\textit{Software Institute -- USI} &
\textit{Univ. Lille, Inria, CNRS, Centrale Lille} \\
\textit{and Lifeware SA} &
\textit{Universit\`a della Svizzera italiana} &
\textit{UMR 9189 CRIStAL} \\
\textit{Graz, Austria} &
\textit{Lugano, Switzerland} &
\textit{Lille, France} \\[0.8em]

Oleksandr Zaitsev &
Robert Peharz &
Romain Robbes \\
oleksandr.zaitsev@cirad.fr &
robert.peharz@tugraz.at &
romain.robbes@labri.fr \\
\textit{CIRAD, UMR SENS} &
\textit{Graz University of Technology} &
\textit{CNRS, University of Bordeaux} \\
\textit{Montpellier, France} &
\textit{Graz, Austria} &
\textit{Bordeaux INP, LaBRI, UMR5800} \\
&
&
\textit{Talence, France} \\[0.9em]

\multicolumn{3}{c}{
\begin{tabular}{>{\centering\arraybackslash}p{0.38\textwidth}
                >{\centering\arraybackslash}p{0.38\textwidth}}

Gabriele Bavota &
St\'ephane Ducasse \\
gabriele.bavota@usi.ch &
stephane.ducasse@inria.fr \\
\textit{Software Institute -- USI} &
\textit{Univ. Lille, Inria, CNRS, Centrale Lille} \\
\textit{Universit\`a della Svizzera italiana} &
\textit{UMR 9189 CRIStAL} \\
\textit{Lugano, Switzerland} &
\textit{Lille, France}

\end{tabular}
} \\[0.4em]

\multicolumn{3}{c}{\footnotesize $^\ast$These authors contributed equally.}

\end{tabular}

\endgroup
}

\maketitle

\begin{abstract}
Large Language Models (LLMs) unlocked new possibilities in automated code writing, becoming the backbone of most code completion tools. While LLMs excel in mainstream languages, they often lack support for the so-called low-resource languages where training data is scarce. As a result, these languages lag behind in the quality of code completion tooling available to their communities. A concrete example is \pharo, a Smalltalk-inspired language whose IDE currently offers only single-token completion. In this work, we report on our experience bringing LLM-based code completion to \pharo. First, we describe an end-to-end pipeline that combines \pharo-specific data curation, continued pre-training and fine-tuning of open code LLMs. Second, we introduce a set of \pharo code completion benchmarks designed to evaluate whether models (i) learn \pharo's syntax and (ii) accurately complete masked Pharo code from real-world GitHub repositories. Third, we show empirically that \pharo-specialized models substantially outperform their original base checkpoints and also exceed the accuracy of substantially larger code LLMs on \pharo completion. Overall, our case study demonstrates the feasibility of bringing strong LLM-based code completion to low-resource programming languages, with models small enough to provide ``real-time'' in-IDE support.
\end{abstract}


\input{introduction}
\input{background}

\input{design}
\input{results}
\input{discussion}

\input{threats}
\input{conclusion}

\section*{Acknowledgments}
Giagnorio and Bavota acknowledge the support of the Swiss National Science Foundation for the PARSED project (grant No. 219294).
AbedelKader, Robbes and Ducasse thank Inria and the LLM4Code challenge for the funding. 
Peharz acknowledge the fundings provided by the Austrian Science Fund (FWF) 10.55776/COE12. The authors also thank Lifeware for providing the computational support required to conduct the experiments in this work.

\bibliographystyle{IEEEtran}
\bibliography{bib}

\end{document}

%% file: introduction.tex
\section{Introduction} \label{sec:intro}
Code completion is a core Integrated Development Environment (IDE) feature that helps developers write code faster, boosts productivity, and cuts manual effort~\cite{Wang23a, Izdai24a}. Early code-completion approaches
typically relied on heuristics~\cite{Robb08a, Hou10a} or machine learning \cite{Bruc09a} to recommend the most probable next token. Early n-gram and neural language models followed \cite{Hind12a, Rayc14a}. However, they struggled to capture broader context and were limited in producing richer, multi-token completions beyond the immediate next token~\cite{Izad22a,Huse25a}.

The advent of Large Language Models (LLMs) has reshaped the landscape of developer tooling, including code completion. LLMs learn statistical regularities from massive source code corpora, and can predict likely continuations of partially written programs, going well beyond next-token suggestions~\cite{Mura24a}. However, the benefits of LLM-based code completion are not evenly distributed across programming languages. Code LLMs \cite{Chen21a, Starcoder, codellama} are usually trained on publicly available repositories (\eg those present on forges like GitHub); this works very well for a prominent language like Python with $\sim$26M public repositories\footnote{All numbers extracted via GitHub advanced search on 18 February 2026.}.
\pharo, the language subject of our case study, has only $\sim$2k public repositories; a full four orders of magnitude less. LLMs performance on coding tasks reflects this imbalance~\cite{Joe25a,Giag25,Cass24a}: they excel on mainstream languages but often struggle with low-resource ones. 

Due to the lack of effective LLM-based code completion tools, \pharo's IDE currently provides only single-token completion \cite{Abed25a}. While valuable, this remains far from the multi-token, context-aware completions users have come to expect in other IDEs. In this work, we introduce effective LLM-based code completion to \pharo and document the engineering and research efforts required to achieve it. \pharo represents an excellent case study for several reasons:  first, it is a (very) low-resource language. To provide a term of comparison, languages considered in the past as ``low-resource'' \cite{Giag25,Cass24a} include Lua (620k public repositories on GitHub), Julia (85k), and Racket (23k), all having at least an order of magnitude more repositories than \pharo (2k). 

Second, \pharo code is saved in a dedicated file format named Tonel (see Figure~\ref{fig:tonel_general_structure}) that supports code import across Smalltalk dialects (\eg Gemstone/S, VA Smalltalk, and VisualWorks). For interoperability reasons, it mixes code and its metadata. In addition, the syntax used for class definition is different from the one used in the IDE or documentation. Such a format may confuse LLMs during training, since the model may learn spurious patterns, misidentify what constitutes executable syntax versus packaging metadata, and blur boundaries between methods and declarations. Third, there are peculiarities of the \pharo syntax inherited from Smalltalk that make it different from mainstream languages, \eg the language control flow operators (\texttt{if}, \texttt{while}) are not part of the syntax but are simple methods. This hinders transfer learning from high-resource languages.

\input{tonel_structure}

Recent work has explored how to improve LLM performance on low-resource languages through different strategies. Some studies focus on data augmentation and corpus construction proposing synthetic data generation or curated multilingual corpora to mitigate data scarcity~\cite{Chen22a,Orla23a,Athi23a,Chai24a}. Complementary efforts investigate cross-lingual transfer and multilingual alignment to better generalize from high-resource languages to underrepresented ones~\cite{Wong25a,Sin25a}. Other approaches rely on continued pre-training and language-specific adaptation techniques, showing measurable improvements through targeted fine-tuning~\cite{Giag25,Joe25a}.

Our goal is, however, in-IDE code completion---rather than offline code generation---that operates under strict latency constraints and must provide incremental, syntactically-valid predictions conditioned on partially written programs~\cite{Svya19a,Svya20a,Svya21a}. This makes the techniques proposed in the literature for code generation (\eg RAG-based solutions) not directly transferable to our setting.  While a few studies examine completion for underrepresented languages~\cite{Gong22a, Popo21a, VanD24a}, none of these address the combined challenges in \pharo: Tonel-aware data curation, extreme data scarcity and real-time IDE integration. To the best of our knowledge, no prior work reports the development of an LLM-based in-IDE completion engine specifically tailored to such a severely low-resource programming language.

We start by building the tools needed for collecting training data from public \pharo projects, customized to deal with the peculiarities of these repositories (\eg Tonel). Then, having the goal of building an efficient completion tool that can run locally on developer machines, we targeted the further pre-training and fine-tuning of ``small'' open code LLMs, namely Qwen2.5 Coder Base \cite{Hui24a} (with model sizes of 0.5B, 1.5B, 3B, and 7B parameters) and Mellum-base \cite{Pavl25a} (4B). These LLMs have been subsequently quantized to ensure fast inference times compatible with what developers perceive as ``real-time'' code completion~\cite{Fran23a,Lin24a,Svya19a,Popo21a,Niel93a}. 

To measure the code completion performance of the trained models, we also built a suite of \pharo code completion benchmarks that evaluate two complementary capabilities: whether a model can learn \pharo's syntax, and whether it can assist developers in realistic coding scenarios. For the former, we (i) translate  the popular HumanEval+ benchmark~\cite{Liu23b} to \pharo and (ii) collect \pharo exercises from Exercism~\cite{Exer26a}, a platform designed to help developers learn programming through coding exercises. These two benchmarks together comprise a total of 211 coding tasks, each featuring a description of the code to implement, a canonical solution, and tests to assess the correctness of proposed solutions. Being in the context of code completion rather than generation, we only exploit the canonical solution and the tests: We randomly mask parts of the canonical solution, ask the LLMs to predict the masked part, and run the obtained code against the tests~\cite{Chen21a,Liu24b}. For the latter, we collect 488 commits performed by developers in public GitHub repositories, identify the lines of code they added, and randomly mask them, asking the models to ``complete the changes''. This scenario represents a simulation of the support that LLMs would have given to developers implementing code changes~\cite{Tufa19a,Chak18a,Zhan23a,Liu24b}. Our empirical results show that \pharo-specialized models significantly outperform the base LLMs and that they also exceed the accuracy of substantially larger ($>$60$\times$) code LLMs. 

In summary, we make the following contributions:
\begin{itemize}
\item \textit{End-to-end specialization pipeline}. We present a practical pipeline for adapting open code LLMs to \pharo statement completion, combining \pharo-specific data curation with continued pre-training and fine-tuning.

\item \textit{Benchmarks for \pharo completion}. We introduce \pharo code completion benchmarks designed to assess (i) syntactic competence in \pharo and (ii) usefulness in realistic coding scenarios.

\item \textit{Empirical evidence of effectiveness}. We demonstrate that \pharo-specialized models outperform their base counterparts and surpass substantially larger code LLMs on \pharo completion, while remaining small enough to support real-time in-IDE usage.
\end{itemize}

%% file: tonel_structure.tex
\begin{figure}[t]
\centering
\begin{lstlisting}[
  basicstyle=\ttfamily\footnotesize,
  backgroundcolor=\color{codegray},
  frame=single
]

    "
    Class comment
    "
    Class {
        #name : 'ClassName',
        #superclass : 'SuperClassName',
        #instVars : [ 'var1', 'var2', ... ],
        #classVars : [ 'default', 'current', ... ],
        #category : 'CategoryName',
        #package : 'PackageName',
        #tag: 'Tag'
        
    }
    
    { #category : 'MethodCategory' }
    ClassName >> methodSelector [
        " Method comment"
        MethodBody
    ]
\end{lstlisting}
\caption{General structure of a Tonel file.}
\label{fig:tonel_general_structure}
\end{figure}

%% file: background.tex
\section{Background and Related Works} \label{sec:brw}

\subsection{\pharo and its completion engine} \label{pharo}
\pharo (\href{https://pharo.org}{pharo.org})~\cite{Blac09a} is an open-source implementation of the Smalltalk programming language~\cite{Gold83} that integrates a dynamic~\cite{Blac09a,Call11a}, reflective~\cite{Thoma24a} language with a live programming environment and tools. It emphasizes simplicity, immediacy, and direct interaction with live objects, enabling developers to modify and inspect running systems without the traditional compile-run cycle~\cite{Kubel18a}.

\subsubsection{The \pharo environment}
\pharo is executed on a dedicated virtual machine (VM)~\cite{Poli14c} as in Java. One of the specific aspects of \pharo (and Smalltalk) is that developers do not explicitly manipulate files. The IDE supports the definition of classes and methods and manages transparently the publication of code either on git (or databases such as ENVY \cite{Pelri01a} for proprietary Smalltalk). In addition, developers can use a docker-like snapshot mechanism called ``image'' that saves/loads frozen memory containing all code entities (packages, classes, methods). This implies that developers do not manually order their code in files, even if they can optionally use a method code formatter. \pharo uses Tonel (\href{https://github.com/pharo-vcs/tonel}{github.com/tonel}) \cite{Duca22c,Poli20y} as a dedicated format to publish code on git. A Tonel file lists a Smalltalk agnostic class definition and its methods alphabetically sorted (see \figref{fig:tonel_general_structure}). 

\subsubsection{\pharo's Syntax}
The \pharo syntax, following that of Smalltalk, has the following characteristics \cite{Zait20a}:

\begin{itemize}
    \item \textbf{Language constructs as messages:} All the language control flow operators (\eg \texttt{if}, \texttt{for}, \texttt{while}) are not part of the syntax but are simple methods.
    \item \textbf{Keyword-based method signatures:} Instead of placing all arguments at the end of a call, arguments are interleaved with parts of the method name, forming readable keyword messages \cite{Spas16a} \eg \lstinline!at: aSymbol ifAbsentPut: aBlock!.
    \item \textbf{English-like, learner-friendly syntax:} Smalltalk's syntax was intentionally designed to be approachable for children~\cite{Kay93a}. The core syntax of \pharo is extremely compact --- it fits on half of a postcard\footnote{\href{https://commons.wikimedia.org/wiki/File:Pharo_syntax_postcard.svg}{wikimedia.org/pharo-syntax-postcard}}. Statements resemble simple English sentences, and periods are used as separators rather than semicolons.
    \item \textbf{Simplified Abstractions:} \pharo introduces streamlined and alternative iterator constructs that reduce complexity.
    \item \textbf{Small methods:} Zaitsev \etal \cite{Zait20a} show that about 50\% of methods feature three or fewer lines of code.

\end{itemize}

\subsubsection{\pharo and code completion}
\paragraph{History} Code completion in \pharo started with a very simple approach~\cite{Koma17a}: the system proposed a list of names matching the typed prefix. Since \pharo is a dynamic language, the search space is large. The system filtered candidates via lightweight parsing and heuristics, such as trying to infer the possible type of a variable by observing the messages previously sent to it \cite{Roma20b,Frol24a}. Although this approach was easy to implement and fast, it often produced noisy suggestions.

A significant step forward was the introduction of semantic completion. Instead of matching only prefixes, the editor parsed the source code, identified the Abstract Syntax Tree (AST) node at the cursor position, and filtered candidates appropriate to that node type. This substantially improved accuracy, but the candidate set could still be too large.

\paragraph{Proposed ranking improvements}
Ranking approaches can help with large candidate lists. Robbes \etal proposed ranking heuristics based on method usage \cite{Robb08a}, and implemented them for Squeak's (\pharo's ancestor) completion engine \cite{Robb10a}.   Romaniuk \etal \cite{Roma20a, Roma20b} proposed ranking candidates using n-gram statistical language models trained on \pharo code. While the n-gram based ranking strategy was proposed and evaluated, it was not integrated into the standard \pharo distribution due to deployment issues.

Oumarou \etal~\cite{Haya22a} proposed a popularity-based ranking mechanism scoring candidates according to usage metrics such as callers, implementors, and references.
However, due to performance concerns, this approach was not integrated into the standard \pharo completion engine.

\pharo has a global namespace for classes, making them particularly sensitive to large candidate lists. AbedelKader \etal~\cite{Abed25a} addressed this by introducing a package-aware heuristic that respects project modularity. The search space has three stages: the current package, the related packages within the same repository, and the global namespace.

\paragraph{\compl} In 2020, \pharo 9 introduced a new completion engine called \compl with a stronger focus on extensibility and responsiveness, thanks to a modular and lazy architecture~\cite{Abed25a}. Prefix filtering, deduplication, and caching mechanisms were added to limit unnecessary computation and improve user interface performance. \compl is the current completion engine in \pharo. Its main limitation is that it recommends a single token at a time.

\subsection{LLMs and low-resource languages}
Recent research has investigated how to improve code generation for low-resource programming languages, highlighting the performance gap that LLMs exhibit between mainstream and underrepresented languages~\cite{Joe25a}. Several studies explicitly quantify this gap and analyze its cause in terms of training data, imbalance, and limited cross-lingual transfer~\cite{Cass24a,Giag25}. A first line of work focuses on language-specific adaptation strategies. Cassano \etal~\cite{Cass24a} introduce MultiPL-T, a framework that generates training data for low-resource programming languages by translating code from high-resource languages such as Python. They show that fine-tuning LLMs on these datasets significantly improves their code generation performance for low-resource languages. Similarly, Giagnorio \etal~\cite{Giag25} study targeted fine-tuning and compare it with in-context learning strategies for low-resource code generation. Fine-tuning consists of further training a pre-trained model on pairs of $\langle$\textit{natural language description}, \textit{code implementation}$\rangle$ written in the target language~\cite{Cass24a}. In-context learning augments the prompt with demonstrations or translations between high-resource and low-resource languages, enabling zero- or few-shot adaptation without weight updates~\cite{Giag25,Athi23a}. Their findings indicate that smaller models benefit from fine-tuning, whereas larger ones are more responsive to prompt-based adaptation.

Beyond direct fine-tuning, alternative approaches have been proposed. Orlandini \etal~\cite{Orla23a} investigate rebalancing multilingual training distributions to mitigate language skew. Dutta \etal~\cite{Dutt24a} apply retrieval-augmented generation to compensate for scarce training data. Paul \etal~\cite{Paul24a} leverage compiler intermediate representations to align code across languages, improving cross-lingual grounding.

Several works extend the evaluation beyond Python-centric benchmarks. Cassano \etal~\cite{Cass22a} introduce multilingual variants of existing code benchmarks, while Athiwaratkun \etal~\cite{Athi23a} and Chai \etal~\cite{Chai24a} construct translated or natively multilingual evaluation suites. Orlandini \etal~\cite{Orla23a} further analyze how benchmark design influences perceived model performance in low-resource settings. Despite these advances, multiple studies consistently 
report persistent performance gaps between high- and low-resource languages~\cite{Giag25,Cass24a}.

Domain-specialized models have been shown to outperform larger general-purpose LLMs when trained on focused corpora. MonoCoder~\cite{Kado24a} and MPIrigen~\cite{Schn24a} achieve superior results in High Performance Computing (HPC) code generation compared to substantially larger models, suggesting that specialization can compensate for scale in some  cases.

Our work differs from the studies above in two main respects. First, we focus on \textit{code completion} rather than offline code generation. Completion operates under strict latency constraints and requires incremental, syntactically-valid predictions within an IDE workflow~\cite{Svya19a,Svya20a}. Second, we target an extremely 
low-resource language (\pharo) characterized by a specific repository structure (Tonel format) and IDE constraints. While prior work has explored completion for underrepresented languages~\cite{Gong22a,VanD24a}, no previous study reports the development of an end-to-end LLM-based completion engine tailored to such a severely low-resource ecosystem.

%% file: design.tex
\section{Study Design} \label{sec:methodology}
Our goal is to investigate the challenges faced when adapting LLMs' code completion capabilities to a low-resource programming language, using \pharo as a case study. We seek to answer the following research question: \emph{To what extent can we enhance the code completion capabilities of LLMs for a low-resource programming language like \pharo?}

Besides answering this question, we also document the suite of tools needed to train and evaluate LLMs on a low-resource language (\ie engineering effort). 

\subsection{Selected Large Language Models}
\label{sec:selection}
We select two families of open-weight models: Qwen2.5 Coder Base \cite{Hui24a} (0.5B, 1.5B, 3B, and 7B parameters) and Mellum-base model \cite{Pavl25a} (4B). Both LLM families have shown great performance on code-related tasks \cite{Aggarwal25a,Pavl25a,Crupi26a}. Importantly, they support fill-in-the-middle (FIM) generation. FIM models can exploit text both before and after the cursor, a capability essential for code completion when editing existing code. The models use different FIM strategies: Qwen2.5 Coder was trained with a prefix-suffix-middle (PSM) objective, while Mellum-base follows a suffix-prefix-middle (SPM) objective. In both cases, the model is trained to predict the missing ``middle'' span of a token sequence given its surrounding context (prefix and suffix). The difference lies in the order in which the model processes the prefix and suffix. 
Considering both strategies gives insights on their influence on the specialization of LLMs to a low-resource language. 

We select a variety of ``small'' model sizes (from 0.5B to 7B), facilitating their deployment on consumer-grade machines. Our largest model (7B), once quantized at 4-bit precision, uses only 4.3GB of memory, making it runnable on recent laptops. We will show that 4-bit quantization does not negatively affect code completion performance.

We fine-tune these models on a curated \pharo dataset we describe next. We compare their performance against their base versions, and against two larger models: Qwen3 Coder 480B A35B Instruct \cite{Qwen3}, one of the most capable open-weight models for code-related tasks; and Claude Sonnet 4.5 \cite{Claude}, one of the commercial models behind Claude Code \cite{ClaudeCode}.

\subsection{Training Datasets}
\label{sec:datasets}
We present here a step-by-step description of how we collected, filtered, and prepared the training dataset used for ``teaching'' \pharo to the selected LLMs.

\subsubsection{Data Collection and Filtering} Involved several filtering steps to ensure data quality and avoid data contamination. 
\label{sub:collection}

\paragraph{Licensing} We mined all GitHub repositories that contain the ``\pharo'' topic and are licensed under the MIT License. We select only MIT-Licensed repositories\footnote{\href{https://opensource.org/license/mit}{opensource.org/license/mit}} because it is a permissive open-source license; this makes it suitable for inclusion in training datasets. Repositories that did not contain an explicit license file were excluded: in the absence of a license, default copyright law applies, and reuse rights are not granted.
This initial search resulted in 748 repositories. 

\paragraph{Version compatibility} To ensure we train our LLMs on current \pharo syntax and APIs, we apply a version-based filter to ensure that the code in the repositories is compatible with \pharo 10 or newer. \pharo 10, released in April 2022\footnote{\href{https://pharo.org/news/pharo10-released.html}{https://pharo.org/news/pharo10-released.html}}, is a stable version of the language; no significant changes were introduced so far. To automate the compatibility validation, we systematically imported each repository into the five newest \pharo environments (\pharo 10 to \pharo 14). If a repository could not be successfully loaded into at least one of these five environments, it was discarded from the dataset. All scripts utilized for this verification step can be found in our replication package~\cite{replication}. Lastly, we exclude all repositories that are not stored in Tonel format \cite{Duca22c,Poli20y}. These filtering steps reduced the number of repositories from 748 to 415.

\paragraph{Splitting and data contamination} Finally, we split the repositories into two sets. We use repositories created before June 1, 2024, for training, and repositories created after this date for repository-level evaluation (see \secref{sub:repository_evaluation}). We use a cut-off date to minimize the chances that the code used for evaluation was seen during pre-training of the selected LLMs. Qwen2.5 Coder was released in September 2024, and has a knowledge cut-off date (\ie the latest date in which training data was collected) of February 2024 \cite{Hui24a}, while Mellum was released in October 2024, but its cut-off date is unclear; this makes June 1st a reasonable date to mitigate data contamination, although we cannot guarantee it. Notably, we make no guarantees for the baselines, Qwen3 Coder, and Claude 4.5 Sonnet. To ensure no contamination of the syntactic evaluation benchmarks (see \secref{sec:syntax-benchs}), we removed from the training dataset all methods that have overlapping 8-grams with the canonical solutions of the tasks in those two benchmarks, as done in previous works \cite{Muennighoff2025a, openr1}.

\subsubsection{Training Instances Construction}
\label{sub:training}
We process the filtered repositories to create two training datasets. We use a \emph{pre-training} dataset to teach LLMs the syntax, structure, and main features of \pharo code, and a \emph{fine-tuning} dataset to align the LLMs to realistic code completion scenarios. 

We implemented two key tools to extract \pharo code from repositories. First, a \pharo lexer in the Pygments library~\cite{pygments}, to tokenize \pharo source code in meaningful units. Second, we integrated the \pharo grammar in the tree-sitter library \cite{treesitter}, to parse \pharo methods and code from the Tonel format to extract its Abstract Syntax Trees (ASTs). Both tools are available in our replication package \cite{replication} to support future research on \pharo code analysis.
Using our parser, we extracted a total of 387{,}159 \pharo methods from the 415 subject repositories.

\paragraph{Pre-training} Following prior work on code completion~\cite{Bava22}, we train on \pharo methods using the left-to-right causal language modeling (CLM) pre-training objective in two different scenarios. For 25\% of the 387{,}159 \pharo methods we mined, we present the entire method to the model and ask it to autoregressively predict each token based only on the tokens to its left. For the remaining 75\%, we remove from each method a contiguous span (``middle'') and train the model to generate the masked span, conditioned on both the prefix and the suffix of the method. The missing span is still generated autoregressively, token by token, but the model's conditioning context includes right-hand information. We use the \pharo parser and lexer to perform an \emph{AST-aware} masking strategy: for each method, we randomly mask an AST node containing between 3 and 10 tokens. The 3-10 token constraint avoids trivial gaps while capping the complexity of the infilling task. We serialize these instances using the model-specific templates: prefix–suffix–middle for Qwen and suffix–prefix–middle for Mellum. The different proportions assigned to the two pre-training settings (\ie full left-to-right method completion versus masked-span completion) are motivated by prior evidence that increasing the share of fill-in-the-middle training instances improves code-completion \cite{Bava22}.

\paragraph{Fine-tuning} For the second training step, we mask the method body with a different strategy, which we call \emph{Random-AST}, similar to the one used for fine-tuning Mellum \cite{Pavl25a}. The \emph{AST-aware} strategy masks well-formed syntactic units to teach structural regularities. However, developers request suggestions at arbitrary cursor positions rather than at pre-selected AST boundaries. \emph{Random-AST} matches a more realistic code-completion usage. Concretely, for each method body we randomly select a token (or token fragment) as the start of the masked code, and extend the mask until the end of the enclosing \textit{statement} AST node. This results in completion targets beginning at unpredictable locations while still considering the syntactic structure.
To preserve the AST-aware knowledge acquired during pre-training, we apply rehearsal training by mixing in a random 20\% sample of instances from the AST-aware dataset into the fine-tuning data \cite{Scialom22a}. The final training dataset for fine-tuning contains 324,725 instances.

\input{benchmarks}

\subsection{Syntax Evaluation Benchmarks}
\label{sec:syntax-benchs}

We created two method-level benchmarks to evaluate LLMs' understanding of \pharo syntax by (i) translating the popular code generation benchmark \ie HumanEval+ \cite{Liu23b} to \pharo, and (ii) collecting \pharo programming exercises from the Exercism platform \cite{Exer26a}. Each benchmark is then divided into two tasks: \emph{AST-Aware} completion and \emph{Random-AST} completion. A tool evaluates the correctness of the results.

\subsubsection{HumanEval+}
\label{sub:syntax_evaluation}
HumanEval+ is a function-level code generation benchmark in which the model must synthesize a function body from a given signature and docstring. It builds on the original HumanEval \cite{Chen21a} suite by providing roughly 80$\times$ additional test cases per task. HumanEval+ also provides a reference (correct) solution for each of its 164 tasks. We select HumanEval+ over more complex alternatives (see \eg \cite{Zhuo2025a, Jimenez2024a}) since it features generic programming tasks that have no dependencies on external libraries or extra context, removing confounding factors such as understanding project-specific code or APIs. While HumanEval+ is a code generation benchmark, we convert it to code completion. We mask spans in the reference implementation, ask the model to fill them in, and then validate the resulting code against the provided test suite. As HumanEval+ is written in Python, we obtained an initial automated translation to \pharo using GPT-4o. However, since GPT-4o has limited knowledge about \pharo, we prompted it with a brief explanation of \pharo's basic syntax and coding conventions along with a list of manually crafted translation examples (few-shot). The prompt also includes selected API mappings from Python to \pharo (\eg \texttt{list} $\rightarrow$ \texttt{OrderedCollection}, \texttt{dictionary} $\rightarrow$ \texttt{Dictionary}, \texttt{None} $\rightarrow$ \texttt{nil}) to help translating the functions' docstring (see full prompt in \cite{replication}). Then, the second author (experienced in code generation benchmark translation) manually checked each translated task and fixed any eventual errors in the docstring, canonical solution, and test cases. The first author (who has three years of working experience in Smalltalk) double-checked the resulting \pharo benchmark. We also checked that the translated code compiles without errors in \pharo 13 (the current stable release), and that the translated solution pass all the translated test cases, fixing any errors we identify. In total, we obtained 164 \pharo tasks from the original Python benchmark.

\subsubsection{Exercism}
For the second benchmark, we collected all 50 programming exercises from the Exercism \pharo track \footnote{https://exercism.org/tracks/pharo-smalltalk} with the exception of trivial ones (\eg \textit{Hello World}). We obtained 47 exercises that we processed to make them ``self-contained''. In particular, to mirror the function-level granularity of HumanEval+, we made sure that each exercise was associated with a single \pharo method representing the reference solution. This means that, if  a reference solution was implemented as a cooperation between multiple methods, we applied an inline method(s) refactoring to obtain a single method dealing with the whole implementation. We then applied the same masking procedure described for HumanEval+ to each reference solution, resulting in the benchmarks summarized in \tabref{tab:benchmarks}.

\subsubsection{Task creation} We processed the HumanEval+ and Exercism instances to create two code completion tasks, using the same masking strategies and settings described in Section \ref{sub:training}. AST-aware masking evaluates a model's capabilities to complete valid AST nodes, while Random-AST masking evaluates a model's performance when random spans of code are masked. \tabref{tab:benchmarks} summarizes the tasks in each benchmark.

\subsubsection{Task verification} To ensure the results are correct, we run the tests. We designed a tool to automatically execute the auto-completed code in a \pharo 13 image and report the test results (\ie pass/fail plus the error messages and stack trace in case of failure). This tool is publicly available \cite{replication}; it is extensible to additional benchmarks to allow future research on \pharo code completion and generation.

\subsection{Repository-level benchmark}
\label{sub:repository_evaluation}
\subsubsection{Instance construction}
To evaluate the performance of our models in more realistic code completion scenarios, we define tasks that emulate real changes implemented by developers during the change history of \pharo systems in our repository-level benchmark.
We mine commits from the 22 test repositories collected as described in Section \ref{sub:collection}. For each commit in the change history, we extract the AST nodes newly introduced by developers (\ie part of added code lines) in all impacted methods, select one of them, and mask a randomly chosen span of 3 to 10 tokens from the node's trailing portion. To increase diversity, we mask each impacted method up to three times (assuming it has at least three new AST nodes).

\subsubsection{Context Modeling}
A key factor in repository-level completion is the amount and relevance of repository context provided to the model in the input prompt. With too little context, the model may not have enough information about the project to support the completion; with too much, the generation speed slows down, and the model may actually get lost in spotting the relevant parts \cite{liu2024lost}. Let $m$ denote the target method containing the masked span, and $ctxt$ denote the additional repository context prepended to the prompt, we experiment with four types of contexts:

\begin{enumerate}
    \item \textbf{No context:} The model receives only the masked target method $m$.
    
    \item \textbf{Class signature:} $ctxt$ includes the declaration of $class(m)$ (\ie the class featuring $m$) and the method signatures (selectors with argument structure) declared in that class, without method bodies.
    
    \item \textbf{Package signature:} $ctxt$ contains the class signatures and method signatures for all classes in the package of $\textit{class}(m)$, without method bodies. 

    \item \textbf{Impacted methods:} $ctxt$ includes the full body of other methods modified within the same commit as $m$, excluding $m$ itself. Note that the order in which the impacted methods appear in the context is random, as the exact order in which the methods were modified is unknown.
\end{enumerate}

Since excessively long inputs may degrade model performance~\cite{Rando25a}, and Mellum exhibits reduced effectiveness beyond 8k tokens~\cite{Pavl25a}, we cap the total context size at 8{,}000 tokens, excluding completion tasks in which at least one of the four contexts exceeded such a threshold. Overall, this procedure resulted in 2{,}185 repository-level completion tasks evaluated under each context policy (see Table \ref{tab:benchmarks}).

\subsection{Experimental Procedure \& Data Analysis}
\subsubsection{Training} We specialize the subject LLMs on the training datasets described in~\secref{sub:training}, following a two-step training procedure. We begin with a pre-training step aimed at teaching LLMs the syntax and APIs of \pharo. Then, we continued with a fine-tuning phase to enhance models on realistic code completion scenarios. 
We use the \textit{LoRA} technique \cite{Hu22a} in both training steps, setting \textit{alpha} to 32, \textit{r} to 16, and \textit{dropout} to 0.05 as done in previous work \cite{Weyssow23a}. We trained the models on a maximum sequence length of 2,048 tokens, with a learning rate of $5 \times 10^{-5}$ and using the AdamW optimizer \cite{Loshchilov19a} with a linear scheduler. Each model has been trained until loss convergence, which typically happened after 3 epochs. For consistency across all models, we always pick the third epoch for our experiments.

\subsubsection{Baselines} We evaluate the trained LLMs across two dimensions: \pharo syntax comprehension and repository-level completion. We compare their performance against (i) their base versions (\ie the same LLMs before the further pre-training and fine-tuning we performed) and (ii) two substantially larger LLMs, namely Qwen3 Coder 480B A35B Instruct \cite{Qwen3} and Claude Sonnet 4.5 \cite{Claude}.

\subsubsection{Syntactic evaluation and metrics} We test each LLM on the four benchmark variants described in \secref{sub:syntax_evaluation}. We collect model predictions by setting the temperature to 0.2; we repeat this process 20 times for each task to account for the stochastic nature of LLMs. Next, we test the LLM-generated completions against the related test suite. We use $pass@k$, as defined by Chen \etal \cite{Chen21a}, as a proxy for the LLMs' performance. This metric evaluates the probability of an LLM to produce a correct solution for a task when $k$ attempts are provided. We report the $pass@1$ across the 20 repetitions of our experiments, \ie we evaluate models' capabilities to provide a correct answer in a single attempt. 
We run McNemar's statistical test \cite{Mcnemar1947a} to determine if there are significant differences in performance between our trained models and their original versions. Since we run multiple comparisons (\eg we compare the pre-trained and fine-tuned versions of each model against the base version), we apply the Benjamini-Hochberg method \cite{Benjamini1995a} to adjust the p-values after the tests. We also report the Odds Ratio (OR) as an effect size of the observed differences.

\subsubsection{Repository-level evaluation and metrics}
Unlike the previous evaluation, the four repository-level benchmark variants presented in~\secref{sub:repository_evaluation} have no associated test suite as they are collected from GitHub commits. Therefore, we evaluate the semantic similarity of the generated code with the original developer-written code. We use two metrics: ChrF \cite{Popovic15a} and CrystalBLEU \cite{Eghbali22a}. ChrF compares the generated output against a reference by measuring overlap in character n-grams. ChrF computes an F-score (harmonic mean) of precision and recall over character n-grams between the generated and reference code. Following previous work \cite{Evtikhiev23a}, we set the beta parameter to 2 and the n-gram order to 6. CrystalBLEU is an adaptation of BLEU \cite{Papineni02a} to code: it removes trivial n-grams (\eg language-specific tokens) from the comparisons. We chose these two metrics as they capture different aspects of the generated code: ChrF measures its closeness to the reference completion on a fine-grained level; CrystalBLEU evaluates the model's performance when predicting meaningful tokens, such as repository-specific identifiers and APIs. Differences across techniques and models were evaluated using the Wilcoxon signed-rank test \cite{wilcoxon} for the statistical significance and the
paired Cliff’s delta \cite{lawrence2005} for the effect size.

%% file: benchmarks.tex
\begin{table}[h]
    \centering
    \footnotesize
    \caption{List of benchmarks used in our experiments}
    \label{tab:benchmarks}
    \relsize{1}
    \begin{tabular}{|l|l|r|}
        \hline
        \rowcolor{black}
        \multicolumn{1}{l!{\color{white}\vrule width 1pt}}{\cellcolor{black}} & \multicolumn{1}{l!{\color{white}\vrule width 1pt}}{\cellcolor{black}\textcolor{white}{\textbf{Benchmark}}} &
        \textcolor{white}{\textbf{\# FIM tasks}} \\ \hline
        
        \multirow{4}{*}{\rotatebox[origin=c]{90}{\scriptsize \textbf{Method-level}}} & HumanEval+ AST-aware   &  2{,}274 \\
        & HumanEval+ R-AST    & 990 \\
        & Exercism AST-aware  & 1{,}272 \\
        & Exercism R-AST      & 551 \\ \hline
        
       \multirow{4}{*}{\rotatebox[origin=c]{90}{\scriptsize \textbf{Repo-level}}} & No context             & 2{,}185 \\
        & Class signatures       & 2{,}185 \\ 
        & Package signatures     & 2{,}185 \\ 
        & Impacted methods       & 2{,}185 \\ \hline
    \end{tabular}
\end{table}

%% file: results.tex
\section{Results}
\label{sec:results}
\input{rq1_results}
We discuss our results in terms of: (i) method-level completion, as assessed on HumanEval+ and Exercism; (ii) repository-level completion and influence of the contextual information provided to the LLM, experimented on the benchmark built starting from real code changes; and (iii) inference time, a first-class requirement in code completion.

\subsection{Method-level Completion}

\subsubsection{Reading the results} \tabref{tab:evaluation_diff_results} reports the $pass@1$ scores for the selected LLMs in three variants: their base checkpoint (top-5 rows), the version after continued pre-training (model names ending with \PT), and the models obtained by fine-tuning on top of pre-training (\SFT). We report results on HumanEval+ and Exercism under both masking scenarios, \ie AST-aware and Random(r)-AST. For each trained model, we use a green underline to indicate a statistically-significant increase in $pass@1$ score with respect to the base checkpoint, and a red dashed underline for significant decreases. The  Odds Ratio (OR) of these comparisons complements the statistical analysis. For example, specializing Qwen2.5 Coder 3B on \pharo code with both pre-training and fine-tuning (see \emph{Qwen2.5 Coder 3B - \SFT} in \tabref{tab:evaluation_diff_results}) leads to a statistically significant increase in performance on HumanEval+ AST-aware: $pass@1$ increases from 71.48\% for the base model to 83.73\% for the \SFT version (+12.25\%). The OR=3.81 indicates $\sim$3 times greater odds to generate a correct completion with respect to the base model.
At the bottom of \tabref{tab:evaluation_diff_results}, we report the results of the two large-sized models used as further baselines (\ie Qwen3 Coder 480B Instruct and Claude 4.5 Sonnet). We statistically compare their $pass@1$ against our best performing fine-tuned model (\ie \emph{Qwen2.5 Coder 7B - \SFT}): a green underline on a large-scale model indicates that it significantly outperforms our 7B LLM, whereas a red dashed underline indicates that it performs significantly worse. 

\subsubsection{Improvement over the base models} Our two-step training procedure substantially improves the capabilities of the selected models to deal with \pharo code. Further pre-training already improves performance on the AST-aware benchmarks, but it is not sufficient to obtain strong results on the more realistic r-AST setting; often, it even \emph{hurts} performance in this setting. Supervised fine-tuning closes the gap in the r-AST setting, while preserving the gains obtained on AST-aware completion. Overall, \emph{Qwen2.5 Coder 7B - \SFT} is the best-performing specialized model in our experiments; \emph{Qwen2.5 Coder 3B - \SFT} represents a strong (cheaper) alternative, with competitive $pass@1$ scores across benchmarks despite exhibiting roughly half the size of the 7B model. Finally, the improvements over the base checkpoints are consistent across the two families of models we considered (\ie Mellum and Qwen), but the magnitude varies. 

\subsubsection{Small specialized LLMs vs large general-purpose models} Comparing our fine-tuned models with Qwen3 Coder 480B Instruct and Claude 4.5 Sonnet, we observe that small LLMs can compete with much larger ones, providing similar or more accurate \pharo completions. While the large models have the edge on the AST-aware benchmark, our specialized Qwen2.5 Coder 3B and 7B models score a higher $pass@1$ than large models on the more realistic r-AST benchmarks. This result shows the feasibility of bringing accurate \pharo completion suggestions even when limited computational resources and/or strict inference-time requirements prevent the use of massive LLMs.

\subsubsection{Reasons behind the failures} We also inspected the type of errors made by LLMs when completing \pharo code. On average, when considering all base LLMs together, we found that failed completions are mostly (65.6\%) due to syntax errors, followed by unexpected exceptions (17.9\%) and assertion failures (16.5\%). This confirms their inability to deal with the \pharo syntax. Our two-step training procedure helps to reduce syntax errors by 33\%, on average, across the five LLMs. A concrete example of a syntax error made by the base models but not by our best specialized LLM is \texttt{TripleSumToZero} (HumanEval+ AST-aware benchmark, id humanevalplus-40-10). This task is algorithmically simple: it checks whether three elements in a collection sum to zero. However, correctness depends on restoring the exact parenthesization and respecting \pharo message precedence \footnote{\href{https://rmod-pharo-mooc.lille.inria.fr/MOOC/PharoMOOC/Week2/C019-W2S03-Messages-Precedence.pdf}{pharo-mooc.org/message-precedence}}. The masked region requires the model to reconstruct an exact \pharo expression inside a parenthesized arithmetic test. \emph{Qwen2.5 Coder 7B - SFT} consistently reconstructs the correct fragment ``\texttt{(aCollection at: i)}'', achieving 20/20 passing runs. All base models generate completions that either break the parentheses structure (\eg ``\texttt{aCollection at: i)}'') or alter the intended evaluation order, leading to syntax errors. All failure reasons are documented in our replication package \cite{replication}.

\begin{findbox}
\textbf{\faLightbulbO~Findings:}
Base LLMs struggle to write correct \pharo code due to their lack of syntax knowledge; our training procedure significantly alleviates this issue, enabling smaller LLMs to be competitive with much larger models.
\end{findbox}

\input{rq2_results}
\subsection{Repository-level evaluation}
Moving from method-level to repository-level completion, we now evaluate how to provide relevant context to improve LLMs' performance on real-world completion tasks.

\tabref{tab:evaluation_commit_results} shows the results achieved by each model on the four repository-level benchmarks described in \secref{sub:repository_evaluation}; each provides a different context for the LLM (\ie no context, class signature, package signature, impacted methods). In addition, there is a further ``control'' context named ``random methods'' that we will detail later. Results are reported in terms of average ChrF and CrystalBLEU  between the generated completions and the original code. The top-5 rows represent the performance of the base models, while those featuring model names ending with \SFT are our best-performing LLMs, namely the further pre-trained and fine-tuned models. Finally, the last two rows report the results of the larger general-purpose LLMs used as baselines.

As in the previous section, green and red underlines represent statistically-significant differences. For example, we can observe that providing the last modified methods as context to \emph{Qwen2.5 Coder 7B - SFT} leads to a significant improvement in code similarity, with an increase of +15.91\% for ChrF and +23.03\% for CrystalBLEU, with respect to the \emph{no context} strategy: ChrF increases from 60.05\% to 75.96\%, while CrystalBLEU increases from 35.96\% to 58.99\%. Our replication package also reports effect size measures (Cliff's delta) of those differences. Given the substantial gap in performance between the base and the fine-tuned models, we focus the statistical analysis on the boost in performance provided by the different contexts with respect to the ``no context'' strategy.

Concerning the low performance of the base models, we notice that the majority of solutions they generate tend to include additional tokens beyond the intended completion (\eg unneeded FIM tokens and extra code). This behavior is problematic for two main reasons. First, it slows down the inference as LLMs continue to produce tokens until a maximum amount is reached. Second, it requires extra post-processing procedures to retain only the most relevant tokens, which can be error-prone and model-specific. In \tabref{tab:evaluation_commit_results}, we use the unprocessed completions to compute the scores, 
while the results after post-processing are available in our replication package \cite{replication}. The main finding (\ie the superiority of the fine-tuned models) does not change by considering the post-processed predictions, but the performance gap shrinks. Still, the additional post-processing for base models is not practical for code completion, particularly in terms of latency.

Shifting the focus to the context strategies, we notice that providing the signatures of class and package methods improves code similarity  on all large-size and fine-tuned models, with Mellum being the only exception. However, this improvement is very limited and often non-statistically significant. 
Providing the last modified methods as context to the same models, instead, significantly boosts ChrF and CrystalBLEU by 14.54\% and 20.21\%, respectively.

To understand whether this improvement is due to the extra \pharo code provided or to the relevance of the provided context, we introduce a ``control'' context treatment named \textit{random methods}. For each task of the benchmark, we select $n$ random methods from the entire repository with $n$ being the same number of methods provided in the \textit{impacted methods} context. We notice that the \textit{random methods} strategy outperforms the \textit{class / package signatures} ones in 37 out of 48 cases, but it is always less effective than the \textit{impacted methods} strategy. From these results, we can conclude that (i) signatures alone are not sufficient to improve repository-level performance, while (ii) relevant method bodies can effectively aid models in generating more accurate completions. Indeed, using the last modified methods as context is the most effective strategy for all models, with average similarity improvements of 10.83\% for ChrF and 12.37\% for CrystalBLEU. The \textit{impacted methods} strategy can be ``replicated'' in real in-IDE completion by tracking the latest methods changed by the developer, using them as contextual information.

Finally, comparing the results of our fine-tuned models with the large-sized baselines, we observe that even the small \emph{Qwen2.5 Coder 1.5B - SFT} with \textit{impacted methods} as context can achieve better performance than the large Qwen3 Coder 480B A35B Instruct, which features $\sim$320$\times$ more parameters. The effectiveness of our specialized models is even more evident when looking at the 3B and 7B versions, which outperform the large Qwen3 Coder 480B A35B Instruct by +7.52\% for ChrF and +8.63\% for CrystalBLEU on average, with the \textit{impacted methods} strategy. 

Claude 4.5 Sonnet remains the best performing LLM, achieving a ChrF score of 83.02\% and CrystalBLEU of 70.52\%.  However, since its training data is private and no information about the knowledge cut-off date has been released, we cannot exclude the risk of data contamination with our benchmark. Indeed, this model achieves high similarity scores even when no context is provided, suggesting that it may be familiar with the repositories we are using for testing.

\begin{findbox}
\textbf{\faLightbulbO~Findings:}
Base models lag behind on repository-level completion tasks, even with additional context. In contrast, our specialized models experience a significant performance boost when provided with the most recently updated methods, surpassing LLMs hundreds of times their size.
\end{findbox}

\subsection{Analysis of Models Latency}
We finally assess the practicality of our best 3B and 7B models for  deployment on consumer hardware. For the 7B model, we applied Q4\_K\_M quantization using llama.cpp~\cite{sparrenberg2025}. We chose this specific quantization approach based on previous work demonstrating its ability to reduce memory footprints while preserving  performance~\cite{Nyamsuren25a}. Quantization reduces the 7B model's memory footprint by approximately 70\% (from 14.19 GiB to 4.36 GiB), with no major loss in performance. Indeed, on the method-level evaluation, the pass@1 metric decreased by 0.61\%, on average, across all benchmarks. On the repository-level benchmark, no changes were observed when considering the best-performing configuration exploiting \textit{impacted methods} as context (ChrF and CrystalBLEU even slightly increased). 

\input{latency}

We computed the average generation (inference) time required by these locally-deployed models when executing the repository-level benchmark with \emph{impacted methods} as context for 10 iterations on different processors (Apple M3 Max and M4 Max) and on a gaming GPU that can be found in consumer-grade hardware (RX 7800XT). As shown in Table~\ref{tab:inference_latency}, the non-quantized 3B model achieves lower inference latency than the quantized 7B model. The 7B model requires 1.3 seconds on average when run on CPUs, which is slightly above the sub-second latency typically desired for interactive code completion~\cite{Mura24a}. Nevertheless, this result suggests that specialized models of this scale are already close to practical in-IDE deployment, while smaller models offer a more favorable latency–accuracy trade-off.

\begin{findbox}
\textbf{\faLightbulbO~Findings:}
Small specialized models can be run on consumer-grade hardware with reasonable inference speed. This makes them practical candidates for interactive \pharo code completion, especially when deployment constraints limit the use of very large models.
\end{findbox}

%% file: rq1_results.tex
\definecolor{veryLightGray}{RGB}{248,248,248}
\definecolor{midGray}{RGB}{220,220,220}

\renewcommand{\tabularxcolumn}[1]{m{#1}}
\newcolumntype{P}{>{\hsize=1.45\hsize\centering\arraybackslash}X}
\newcolumntype{O}{>{\hsize=0.55\hsize\centering\arraybackslash}X}

\renewcommand{\ULthickness}{1pt}
\renewcommand{\ULdepth}{3.2pt}

\newcommand{\thickdash}[1]{
     \begin{tikzpicture}[baseline=(toset.base)]
         \node[inner sep=0pt, outer sep=0pt, anchor=base] (toset) {#1};
         \draw[line width=1pt, dash pattern=on 3.5pt off 2pt, dash phase=1.4pt] ([yshift=-2.5pt]toset.base west) -- ([yshift=-2.5pt]toset.base east);
     \end{tikzpicture}
}

\begin{table*}[ht!]
    \caption{Method-level completion: Pass@1 comparison between base, pre-trained, and fine-tuned models.}
    \label{tab:evaluation_diff_results}
    \footnotesize
    \begin{tcolorbox}[
        boxrule=0.5pt,
        colback=white,
        colframe=black!80,
        sharp corners,
        boxsep=0pt, left=0pt, right=0pt, top=0pt, bottom=0pt]
    \relsize{-1}
    \renewcommand{\arraystretch}{1.3}
    \setlength{\tabcolsep}{2pt}
    \begin{tabularx}{\linewidth}{l!{\vrule width 1pt}P|O|P|O!{\vrule width 1pt}P|O|P|O}
        \rowcolor{black}
        \multicolumn{1}{l!{\color{white}\vrule width 1pt}}{\cellcolor{black}} & \multicolumn{4}{c!{\color{white}\vrule width 1pt}}{\textcolor{white}{\textbf{HumanEval+}}} & \multicolumn{4}{c}{\textcolor{white}{\textbf{Exercism}}} \\
        \Xhline{1\arrayrulewidth}
        \rowcolor{midGray}
        \cellcolor{veryLightGray} & \multicolumn{2}{c|}{\textbf{AST-aware}} & \multicolumn{2}{c!{\vrule width 1pt}}{\textbf{r-AST}} & \multicolumn{2}{c|}{\textbf{AST-aware}} & \multicolumn{2}{c}{\textbf{r-AST}} \\
        \hhline{>{\arrayrulecolor{veryLightGray}}->{\arrayrulecolor{black}}!{\vrule width 1pt}----!{\vrule width 1pt}----}
        \rowcolor{veryLightGray}
        \multirow{-2}{*}{\textbf{Model}} & \textbf{Pass@1} & \textbf{OR} & \textbf{Pass@1} & \textbf{OR} & \textbf{Pass@1} & \textbf{OR} & \textbf{Pass@1} & \textbf{OR} \\
        \Xhline{2\arrayrulewidth}
        Mellum 4B Base & \mbox{\makebox[0.5em]{}\makebox[2.5em][r]{59.41}\hspace{0.2em}\makebox[3.5em]{}} & --- & \mbox{\makebox[0.5em]{}\makebox[2.5em][r]{35.02}\hspace{0.2em}\makebox[3.5em]{}} & --- & \mbox{\makebox[0.5em]{}\makebox[2.5em][r]{52.24}\hspace{0.2em}\makebox[3.5em]{}} & --- & \mbox{\makebox[0.5em]{}\makebox[2.5em][r]{26.62}\hspace{0.2em}\makebox[3.5em]{}} & --- \\
        Qwen 2.5 Coder 0.5B & \mbox{\makebox[0.5em]{}\makebox[2.5em][r]{32.52}\hspace{0.2em}\makebox[3.5em]{}} & --- & \mbox{\makebox[0.5em]{}\makebox[2.5em][r]{18.14}\hspace{0.2em}\makebox[3.5em]{}} & --- & \mbox{\makebox[0.5em]{}\makebox[2.5em][r]{20.43}\hspace{0.2em}\makebox[3.5em]{}} & --- & \mbox{\makebox[0.5em]{}\makebox[2.5em][r]{\hphantom{0}9.12}\hspace{0.2em}\makebox[3.5em]{}} & --- \\
        Qwen 2.5 Coder 1.5B & \mbox{\makebox[0.5em]{}\makebox[2.5em][r]{29.77}\hspace{0.2em}\makebox[3.5em]{}} & --- & \mbox{\makebox[0.5em]{}\makebox[2.5em][r]{11.97}\hspace{0.2em}\makebox[3.5em]{}} & --- & \mbox{\makebox[0.5em]{}\makebox[2.5em][r]{17.67}\hspace{0.2em}\makebox[3.5em]{}} & --- & \mbox{\makebox[0.5em]{}\makebox[2.5em][r]{\hphantom{0}5.25}\hspace{0.2em}\makebox[3.5em]{}} & --- \\
        Qwen 2.5 Coder 3B & \mbox{\makebox[0.5em]{}\makebox[2.5em][r]{71.48}\hspace{0.2em}\makebox[3.5em]{}} & --- & \mbox{\makebox[0.5em]{}\makebox[2.5em][r]{44.47}\hspace{0.2em}\makebox[3.5em]{}} & --- & \mbox{\makebox[0.5em]{}\makebox[2.5em][r]{64.85}\hspace{0.2em}\makebox[3.5em]{}} & --- & \mbox{\makebox[0.5em]{}\makebox[2.5em][r]{31.81}\hspace{0.2em}\makebox[3.5em]{}} & --- \\
        Qwen 2.5 Coder 7B & \mbox{\makebox[0.5em]{}\makebox[2.5em][r]{71.12}\hspace{0.2em}\makebox[3.5em]{}} & --- & \mbox{\makebox[0.5em]{}\makebox[2.5em][r]{45.24}\hspace{0.2em}\makebox[3.5em]{}} & --- & \mbox{\makebox[0.5em]{}\makebox[2.5em][r]{68.58}\hspace{0.2em}\makebox[3.5em]{}} & --- & \mbox{\makebox[0.5em]{}\makebox[2.5em][r]{36.12}\hspace{0.2em}\makebox[3.5em]{}} & --- \\[.5ex]
        \Xhline{2\arrayrulewidth}
        Mellum 4B Base - PT & \mbox{\makebox[0.5em]{}\makebox[2.5em][r]{72.78}\hspace{0.2em}\makebox[3.5em][l]{\textcolor{ForestGreen}{\uline{\textcolor{black}{{\tiny(\makebox[1.2ex][c]{+}13.37)}}}}}} & \hphantom{0}3.19 & \mbox{\makebox[0.5em]{}\makebox[2.5em][r]{33.14}\hspace{0.2em}\makebox[3.5em][l]{\textcolor{BrickRed}{\thickdash{\textcolor{black}{{\tiny(\makebox[1.2ex][c]{-}\hphantom{0}1.88)}}}}}} & \hphantom{0}0.84 & \mbox{\makebox[0.5em]{}\makebox[2.5em][r]{71.70}\hspace{0.2em}\makebox[3.5em][l]{\textcolor{ForestGreen}{\uline{\textcolor{black}{{\tiny(\makebox[1.2ex][c]{+}19.46)}}}}}} & \hphantom{0}5.84 & \mbox{\makebox[0.5em]{}\makebox[2.5em][r]{24.74}\hspace{0.2em}\makebox[3.5em][l]{\textcolor{BrickRed}{\thickdash{\textcolor{black}{{\tiny(\makebox[1.2ex][c]{-}\hphantom{0}1.88)}}}}}} & \hphantom{0}0.81 \\
        Qwen 2.5 Coder 0.5B - PT & \mbox{\makebox[0.5em]{}\makebox[2.5em][r]{61.08}\hspace{0.2em}\makebox[3.5em][l]{\textcolor{ForestGreen}{\uline{\textcolor{black}{{\tiny(\makebox[1.2ex][c]{+}28.56)}}}}}} & \hphantom{0}8.06 & \mbox{\makebox[0.5em]{}\makebox[2.5em][r]{13.92}\hspace{0.2em}\makebox[3.5em][l]{\textcolor{BrickRed}{\thickdash{\textcolor{black}{{\tiny(\makebox[1.2ex][c]{-}\hphantom{0}4.22)}}}}}} & \hphantom{0}0.63 & \mbox{\makebox[0.5em]{}\makebox[2.5em][r]{51.15}\hspace{0.2em}\makebox[3.5em][l]{\textcolor{ForestGreen}{\uline{\textcolor{black}{{\tiny(\makebox[1.2ex][c]{+}30.72)}}}}}} & 11.59 & \mbox{\makebox[0.5em]{}\makebox[2.5em][r]{\hphantom{0}9.83}\hspace{0.2em}\makebox[3.5em][l]{\textcolor{ForestGreen}{\uline{\textcolor{black}{{\tiny(\makebox[1.2ex][c]{+}\hphantom{0}0.71)}}}}}} & \hphantom{0}1.12 \\
        Qwen 2.5 Coder 1.5B - PT & \mbox{\makebox[0.5em]{}\makebox[2.5em][r]{79.21}\hspace{0.2em}\makebox[3.5em][l]{\textcolor{ForestGreen}{\uline{\textcolor{black}{{\tiny(\makebox[1.2ex][c]{+}49.44)}}}}}} & 20.62 & \mbox{\makebox[0.5em]{}\makebox[2.5em][r]{21.16}\hspace{0.2em}\makebox[3.5em][l]{\textcolor{ForestGreen}{\uline{\textcolor{black}{{\tiny(\makebox[1.2ex][c]{+}\hphantom{0}9.19)}}}}}} & \hphantom{0}2.22 & \mbox{\makebox[0.5em]{}\makebox[2.5em][r]{71.51}\hspace{0.2em}\makebox[3.5em][l]{\textcolor{ForestGreen}{\uline{\textcolor{black}{{\tiny(\makebox[1.2ex][c]{+}53.84)}}}}}} & 35.59 & \mbox{\makebox[0.5em]{}\makebox[2.5em][r]{15.90}\hspace{0.2em}\makebox[3.5em][l]{\textcolor{ForestGreen}{\uline{\textcolor{black}{{\tiny(\makebox[1.2ex][c]{+}10.65)}}}}}} & \hphantom{0}3.88 \\
        Qwen 2.5 Coder 3B - PT & \mbox{\makebox[0.5em]{}\makebox[2.5em][r]{84.19}\hspace{0.2em}\makebox[3.5em][l]{\textcolor{ForestGreen}{\uline{\textcolor{black}{{\tiny(\makebox[1.2ex][c]{+}12.71)}}}}}} & \hphantom{0}4.23 & \mbox{\makebox[0.5em]{}\makebox[2.5em][r]{25.26}\hspace{0.2em}\makebox[3.5em][l]{\textcolor{BrickRed}{\thickdash{\textcolor{black}{{\tiny(\makebox[1.2ex][c]{-}19.21)}}}}}} & \hphantom{0}0.20 & \mbox{\makebox[0.5em]{}\makebox[2.5em][r]{77.36}\hspace{0.2em}\makebox[3.5em][l]{\textcolor{ForestGreen}{\uline{\textcolor{black}{{\tiny(\makebox[1.2ex][c]{+}12.51)}}}}}} & \hphantom{0}4.07 & \mbox{\makebox[0.5em]{}\makebox[2.5em][r]{19.49}\hspace{0.2em}\makebox[3.5em][l]{\textcolor{BrickRed}{\thickdash{\textcolor{black}{{\tiny(\makebox[1.2ex][c]{-}12.32)}}}}}} & \hphantom{0}0.28 \\
        Qwen 2.5 Coder 7B - PT & \mbox{\makebox[0.5em]{}\makebox[2.5em][r]{89.11}\hspace{0.2em}\makebox[3.5em][l]{\textcolor{ForestGreen}{\uline{\textcolor{black}{{\tiny(\makebox[1.2ex][c]{+}17.99)}}}}}} & \hphantom{0}7.09 & \mbox{\makebox[0.5em]{}\makebox[2.5em][r]{27.65}\hspace{0.2em}\makebox[3.5em][l]{\textcolor{BrickRed}{\thickdash{\textcolor{black}{{\tiny(\makebox[1.2ex][c]{-}17.59)}}}}}} & \hphantom{0}0.26 & \mbox{\makebox[0.5em]{}\makebox[2.5em][r]{84.75}\hspace{0.2em}\makebox[3.5em][l]{\textcolor{ForestGreen}{\uline{\textcolor{black}{{\tiny(\makebox[1.2ex][c]{+}16.17)}}}}}} & \hphantom{0}6.53 & \mbox{\makebox[0.5em]{}\makebox[2.5em][r]{21.45}\hspace{0.2em}\makebox[3.5em][l]{\textcolor{BrickRed}{\thickdash{\textcolor{black}{{\tiny(\makebox[1.2ex][c]{-}14.67)}}}}}} & \hphantom{0}0.21 \\[.5ex]
        \Xhline{2\arrayrulewidth}
        Mellum 4B Base - SFT & \mbox{\makebox[0.5em]{}\makebox[2.5em][r]{69.81}\hspace{0.2em}\makebox[3.5em][l]{\textcolor{ForestGreen}{\uline{\textcolor{black}{{\tiny(\makebox[1.2ex][c]{+}10.40)}}}}}} & \hphantom{0}2.21 & \mbox{\makebox[0.5em]{}\makebox[2.5em][r]{46.70}\hspace{0.2em}\makebox[3.5em][l]{\textcolor{ForestGreen}{\uline{\textcolor{black}{{\tiny(\makebox[1.2ex][c]{+}11.68)}}}}}} & \hphantom{0}3.77 & \mbox{\makebox[0.5em]{}\makebox[2.5em][r]{66.99}\hspace{0.2em}\makebox[3.5em][l]{\textcolor{ForestGreen}{\uline{\textcolor{black}{{\tiny(\makebox[1.2ex][c]{+}14.75)}}}}}} & \hphantom{0}3.29 & \mbox{\makebox[0.5em]{}\makebox[2.5em][r]{35.34}\hspace{0.2em}\makebox[3.5em][l]{\textcolor{ForestGreen}{\uline{\textcolor{black}{{\tiny(\makebox[1.2ex][c]{+}\hphantom{0}8.72)}}}}}} & \hphantom{0}2.87 \\
        Qwen 2.5 Coder 0.5B - SFT & \mbox{\makebox[0.5em]{}\makebox[2.5em][r]{62.16}\hspace{0.2em}\makebox[3.5em][l]{\textcolor{ForestGreen}{\uline{\textcolor{black}{{\tiny(\makebox[1.2ex][c]{+}29.64)}}}}}} & \hphantom{0}8.02 & \mbox{\makebox[0.5em]{}\makebox[2.5em][r]{37.96}\hspace{0.2em}\makebox[3.5em][l]{\textcolor{ForestGreen}{\uline{\textcolor{black}{{\tiny(\makebox[1.2ex][c]{+}19.82)}}}}}} & \hphantom{0}7.37 & \mbox{\makebox[0.5em]{}\makebox[2.5em][r]{55.61}\hspace{0.2em}\makebox[3.5em][l]{\textcolor{ForestGreen}{\uline{\textcolor{black}{{\tiny(\makebox[1.2ex][c]{+}35.18)}}}}}} & 16.92 & \mbox{\makebox[0.5em]{}\makebox[2.5em][r]{27.01}\hspace{0.2em}\makebox[3.5em][l]{\textcolor{ForestGreen}{\uline{\textcolor{black}{{\tiny(\makebox[1.2ex][c]{+}17.89)}}}}}} & 18.76 \\
        Qwen 2.5 Coder 1.5B - SFT & \mbox{\makebox[0.5em]{}\makebox[2.5em][r]{78.66}\hspace{0.2em}\makebox[3.5em][l]{\textcolor{ForestGreen}{\uline{\textcolor{black}{{\tiny(\makebox[1.2ex][c]{+}48.89)}}}}}} & 19.75 & \mbox{\makebox[0.5em]{}\makebox[2.5em][r]{48.71}\hspace{0.2em}\makebox[3.5em][l]{\textcolor{ForestGreen}{\uline{\textcolor{black}{{\tiny(\makebox[1.2ex][c]{+}36.74)}}}}}} & 28.24 & \mbox{\makebox[0.5em]{}\makebox[2.5em][r]{71.87}\hspace{0.2em}\makebox[3.5em][l]{\textcolor{ForestGreen}{\uline{\textcolor{black}{{\tiny(\makebox[1.2ex][c]{+}54.20)}}}}}} & 37.48 & \mbox{\makebox[0.5em]{}\makebox[2.5em][r]{35.41}\hspace{0.2em}\makebox[3.5em][l]{\textcolor{ForestGreen}{\uline{\textcolor{black}{{\tiny(\makebox[1.2ex][c]{+}30.16)}}}}}} & 28.92 \\
        Qwen 2.5 Coder 3B - SFT & \mbox{\makebox[0.5em]{}\makebox[2.5em][r]{83.73}\hspace{0.2em}\makebox[3.5em][l]{\textcolor{ForestGreen}{\uline{\textcolor{black}{{\tiny(\makebox[1.2ex][c]{+}12.25)}}}}}} & \hphantom{0}3.81 & \mbox{\makebox[0.5em]{}\makebox[2.5em][r]{51.83}\hspace{0.2em}\makebox[3.5em][l]{\textcolor{ForestGreen}{\uline{\textcolor{black}{{\tiny(\makebox[1.2ex][c]{+}\hphantom{0}7.36)}}}}}} & \hphantom{0}3.38 & \mbox{\makebox[0.5em]{}\makebox[2.5em][r]{78.11}\hspace{0.2em}\makebox[3.5em][l]{\textcolor{ForestGreen}{\uline{\textcolor{black}{{\tiny(\makebox[1.2ex][c]{+}13.26)}}}}}} & \hphantom{0}3.92 & \mbox{\makebox[0.5em]{}\makebox[2.5em][r]{41.81}\hspace{0.2em}\makebox[3.5em][l]{\textcolor{ForestGreen}{\uline{\textcolor{black}{{\tiny(\makebox[1.2ex][c]{+}10.00)}}}}}} & \hphantom{0}4.65 \\
        Qwen 2.5 Coder 7B - SFT & \mbox{\makebox[0.5em]{}\makebox[2.5em][r]{89.04}\hspace{0.2em}\makebox[3.5em][l]{\textcolor{ForestGreen}{\uline{\textcolor{black}{{\tiny(\makebox[1.2ex][c]{+}17.92)}}}}}} & \hphantom{0}6.10 & \mbox{\makebox[0.5em]{}\makebox[2.5em][r]{$\textbf{52.76}$}\hspace{0.2em}\makebox[3.5em][l]{\textcolor{ForestGreen}{\uline{\textcolor{black}{{\tiny(\makebox[1.2ex][c]{+}\hphantom{0}7.52)}}}}}} & \hphantom{0}3.21 & \mbox{\makebox[0.5em]{}\makebox[2.5em][r]{85.84}\hspace{0.2em}\makebox[3.5em][l]{\textcolor{ForestGreen}{\uline{\textcolor{black}{{\tiny(\makebox[1.2ex][c]{+}17.26)}}}}}} & \hphantom{0}7.58 & \mbox{\makebox[0.5em]{}\makebox[2.5em][r]{42.95}\hspace{0.2em}\makebox[3.5em][l]{\textcolor{ForestGreen}{\uline{\textcolor{black}{{\tiny(\makebox[1.2ex][c]{+}\hphantom{0}6.83)}}}}}} & \hphantom{0}3.64 \\[.5ex]
        \Xhline{4\arrayrulewidth}
        Qwen 3 Coder 480B A35B Instruct & \mbox{\makebox[0.5em]{}\makebox[2.5em][r]{91.95}\hspace{0.2em}\makebox[3.5em][l]{\textcolor{ForestGreen}{\uline{\textcolor{black}{{\tiny(\makebox[1.2ex][c]{+}\hphantom{0}2.91)}}}}}} & \hphantom{0}1.71 & \mbox{\makebox[0.5em]{}\makebox[2.5em][r]{45.13}\hspace{0.2em}\makebox[3.5em][l]{\textcolor{BrickRed}{\thickdash{\textcolor{black}{{\tiny(\makebox[1.2ex][c]{-}\hphantom{0}7.63)}}}}}} & \hphantom{0}0.39 & \mbox{\makebox[0.5em]{}\makebox[2.5em][r]{90.35}\hspace{0.2em}\makebox[3.5em][l]{\textcolor{ForestGreen}{\uline{\textcolor{black}{{\tiny(\makebox[1.2ex][c]{+}\hphantom{0}4.51)}}}}}} & \hphantom{0}2.08 & \mbox{\makebox[0.5em]{}\makebox[2.5em][r]{36.92}\hspace{0.2em}\makebox[3.5em][l]{\textcolor{BrickRed}{\thickdash{\textcolor{black}{{\tiny(\makebox[1.2ex][c]{-}\hphantom{0}6.03)}}}}}} & \hphantom{0}0.48 \\
        Claude 4.5 Sonnet & \mbox{\makebox[0.5em]{}\makebox[2.5em][r]{$\textbf{95.07}$}\hspace{0.2em}\makebox[3.5em][l]{\textcolor{ForestGreen}{\uline{\textcolor{black}{{\tiny(\makebox[1.2ex][c]{+}\hphantom{0}6.03)}}}}}} & \hphantom{0}3.46 & \mbox{\makebox[0.5em]{}\makebox[2.5em][r]{51.53}\hspace{0.2em}\makebox[3.5em][l]{\textcolor{BrickRed}{\thickdash{\textcolor{black}{{\tiny(\makebox[1.2ex][c]{-}\hphantom{0}1.23)}}}}}} & \hphantom{0}0.79 & \mbox{\makebox[0.5em]{}\makebox[2.5em][r]{$\textbf{91.75}$}\hspace{0.2em}\makebox[3.5em][l]{\textcolor{ForestGreen}{\uline{\textcolor{black}{{\tiny(\makebox[1.2ex][c]{+}\hphantom{0}5.91)}}}}}} & \hphantom{0}2.75 & \mbox{\makebox[0.5em]{}\makebox[2.5em][r]{41.11}\hspace{0.2em}\makebox[3.5em][l]{\textcolor{BrickRed}{\thickdash{\textcolor{black}{{\tiny(\makebox[1.2ex][c]{-}\hphantom{0}1.84)}}}}}} & \hphantom{0}0.78 \\[.5ex]
    \end{tabularx}
    \end{tcolorbox}
\end{table*}

%% file: rq2_results.tex
\definecolor{veryLightGray}{RGB}{248,248,248}
\definecolor{midGray}{RGB}{220,220,220}

\renewcommand{\tabularxcolumn}[1]{m{#1}}
\newcolumntype{F}{>{\hsize=0.975\hsize\centering\arraybackslash}X}
\newcolumntype{B}{>{\hsize=1.025\hsize\centering\arraybackslash}X}

\renewcommand{\ULthickness}{1pt}
\renewcommand{\ULdepth}{2.8pt}

\newcommand{\newthickdash}[1]{
     \begin{tikzpicture}[baseline=(toset.base)]
         \node[inner sep=0pt, outer sep=0pt, anchor=base] (toset) {#1};
         \draw[line width=1pt, dash pattern=on 3.5pt off 2pt, dash phase=1.3pt] ([yshift=-2pt]toset.base west) -- ([yshift=-2pt]toset.base east);
     \end{tikzpicture}
}

\begin{table*}[thb]
    \caption{Repository-level completion: ChrF and CrystalBLEU of generated completions. Higher values indicate better performance.}
    \label{tab:evaluation_commit_results}
    \scriptsize
    \begin{tcolorbox}[
        boxrule=0.5pt,
        colback=white,
        colframe=black!80,
        sharp corners,
        boxsep=0pt, left=0pt, right=0pt, top=0pt, bottom=0pt]
    \relsize{-1}
    \renewcommand{\arraystretch}{1.3}
    \setlength{\tabcolsep}{1pt}
    \begin{tabularx}{\linewidth}{l!{\vrule width 1pt}F|B!{\vrule width 1pt}F|B!{\vrule width 1pt}F|B!{\vrule width 1pt}F|B!{\vrule width 1pt}F|B}
        \rowcolor{black}
        \multicolumn{1}{l!{\color{white}\vrule width 1pt}}{\cellcolor{black}} & \multicolumn{2}{c!{\color{white}\vrule width 1pt}}{\textcolor{white}{\textbf{No Context}}} & \multicolumn{2}{c!{\color{white}\vrule width 1pt}}{\textcolor{white}{\textbf{Class Signatures}}} & \multicolumn{2}{c!{\color{white}\vrule width 1pt}}{\textcolor{white}{\textbf{Package Signatures}}} & \multicolumn{2}{c!{\color{white}\vrule width 1pt}}{\textcolor{white}{\textbf{Impacted Methods}}} & \multicolumn{2}{c}{\textcolor{white}{\textbf{Random Methods}}} \\
        \Xhline{1\arrayrulewidth}
        \rowcolor{veryLightGray}
        \textbf{Model} & \textbf{ChrF} & \textbf{CrystalBLEU} & \textbf{ChrF} & \textbf{CrystalBLEU} & \textbf{ChrF} & \textbf{CrystalBLEU} & \textbf{ChrF} & \textbf{CrystalBLEU} & \textbf{ChrF} & \textbf{CrystalBLEU} \\
        \Xhline{2\arrayrulewidth}
        Mellum 4B Base & \mbox{\makebox[2.3em][r]{52.61}\hspace{0.05em}\makebox[2.7em]{}} & \mbox{\makebox[2.3em][r]{26.46}\hspace{0.05em}\makebox[2.7em]{}} & \mbox{\makebox[2.3em][r]{51.08}\hspace{0.05em}\makebox[2.7em][l]{\textcolor{BrickRed}{\newthickdash{\textcolor{black}{\tiny (\makebox[1.2ex][c]{-}\hphantom{0}1.53)}}}}} & \mbox{\makebox[2.3em][r]{24.04}\hspace{0.05em}\makebox[2.7em][l]{\textcolor{BrickRed}{\newthickdash{\textcolor{black}{\tiny (\makebox[1.2ex][c]{-}\hphantom{0}2.42)}}}}} & \mbox{\makebox[2.3em][r]{46.78}\hspace{0.05em}\makebox[2.7em][l]{\textcolor{BrickRed}{\newthickdash{\textcolor{black}{\tiny (\makebox[1.2ex][c]{-}\hphantom{0}5.83)}}}}} & \mbox{\makebox[2.3em][r]{19.82}\hspace{0.05em}\makebox[2.7em][l]{\textcolor{BrickRed}{\newthickdash{\textcolor{black}{\tiny (\makebox[1.2ex][c]{-}\hphantom{0}6.64)}}}}} & \mbox{\makebox[2.3em][r]{53.41}\hspace{0.05em}\makebox[2.7em][l]{\textcolor{black}{\tiny (\makebox[1.2ex][c]{+}\hphantom{0}0.80)}}} & \mbox{\makebox[2.3em][r]{25.87}\hspace{0.05em}\makebox[2.7em][l]{\textcolor{black}{\tiny (\makebox[1.2ex][c]{-}\hphantom{0}0.59)}}} & \mbox{\makebox[2.3em][r]{46.56}\hspace{0.05em}\makebox[2.7em][l]{\textcolor{BrickRed}{\newthickdash{\textcolor{black}{\tiny (\makebox[1.2ex][c]{-}\hphantom{0}6.05)}}}}} & \mbox{\makebox[2.3em][r]{19.62}\hspace{0.05em}\makebox[2.7em][l]{\textcolor{BrickRed}{\newthickdash{\textcolor{black}{\tiny (\makebox[1.2ex][c]{-}\hphantom{0}6.84)}}}}} \\
        Qwen 2.5 Coder 0.5B & \mbox{\makebox[2.3em][r]{22.15}\hspace{0.05em}\makebox[2.7em]{}} & \mbox{\makebox[2.3em][r]{\hphantom{0}2.16}\hspace{0.05em}\makebox[2.7em]{}} & \mbox{\makebox[2.3em][r]{23.11}\hspace{0.05em}\makebox[2.7em][l]{\textcolor{black}{\tiny (\makebox[1.2ex][c]{+}\hphantom{0}0.96)}}} & \mbox{\makebox[2.3em][r]{\hphantom{0}2.79}\hspace{0.05em}\makebox[2.7em][l]{\textcolor{ForestGreen}{\uline{\textcolor{black}{\tiny (\makebox[1.2ex][c]{+}\hphantom{0}0.63)}}}}} & \mbox{\makebox[2.3em][r]{22.83}\hspace{0.05em}\makebox[2.7em][l]{\textcolor{black}{\tiny (\makebox[1.2ex][c]{+}\hphantom{0}0.68)}}} & \mbox{\makebox[2.3em][r]{\hphantom{0}2.74}\hspace{0.05em}\makebox[2.7em][l]{\textcolor{ForestGreen}{\uline{\textcolor{black}{\tiny (\makebox[1.2ex][c]{+}\hphantom{0}0.58)}}}}} & \mbox{\makebox[2.3em][r]{31.07}\hspace{0.05em}\makebox[2.7em][l]{\textcolor{ForestGreen}{\uline{\textcolor{black}{\tiny (\makebox[1.2ex][c]{+}\hphantom{0}8.92)}}}}} & \mbox{\makebox[2.3em][r]{\hphantom{0}5.44}\hspace{0.05em}\makebox[2.7em][l]{\textcolor{ForestGreen}{\uline{\textcolor{black}{\tiny (\makebox[1.2ex][c]{+}\hphantom{0}3.28)}}}}} & \mbox{\makebox[2.3em][r]{25.07}\hspace{0.05em}\makebox[2.7em][l]{\textcolor{ForestGreen}{\uline{\textcolor{black}{\tiny (\makebox[1.2ex][c]{+}\hphantom{0}2.92)}}}}} & \mbox{\makebox[2.3em][r]{\hphantom{0}2.98}\hspace{0.05em}\makebox[2.7em][l]{\textcolor{ForestGreen}{\uline{\textcolor{black}{\tiny (\makebox[1.2ex][c]{+}\hphantom{0}0.82)}}}}} \\
        Qwen 2.5 Coder 1.5B & \mbox{\makebox[2.3em][r]{28.16}\hspace{0.05em}\makebox[2.7em]{}} & \mbox{\makebox[2.3em][r]{\hphantom{0}4.98}\hspace{0.05em}\makebox[2.7em]{}} & \mbox{\makebox[2.3em][r]{26.15}\hspace{0.05em}\makebox[2.7em][l]{\textcolor{BrickRed}{\newthickdash{\textcolor{black}{\tiny (\makebox[1.2ex][c]{-}\hphantom{0}2.01)}}}}} & \mbox{\makebox[2.3em][r]{\hphantom{0}3.22}\hspace{0.05em}\makebox[2.7em][l]{\textcolor{black}{\tiny (\makebox[1.2ex][c]{-}\hphantom{0}1.76)}}} & \mbox{\makebox[2.3em][r]{26.77}\hspace{0.05em}\makebox[2.7em][l]{\textcolor{BrickRed}{\newthickdash{\textcolor{black}{\tiny (\makebox[1.2ex][c]{-}\hphantom{0}1.39)}}}}} & \mbox{\makebox[2.3em][r]{\hphantom{0}3.40}\hspace{0.05em}\makebox[2.7em][l]{\textcolor{black}{\tiny (\makebox[1.2ex][c]{-}\hphantom{0}1.58)}}} & \mbox{\makebox[2.3em][r]{34.16}\hspace{0.05em}\makebox[2.7em][l]{\textcolor{ForestGreen}{\uline{\textcolor{black}{\tiny (\makebox[1.2ex][c]{+}\hphantom{0}6.00)}}}}} & \mbox{\makebox[2.3em][r]{\hphantom{0}6.44}\hspace{0.05em}\makebox[2.7em][l]{\textcolor{ForestGreen}{\uline{\textcolor{black}{\tiny (\makebox[1.2ex][c]{+}\hphantom{0}1.46)}}}}} & \mbox{\makebox[2.3em][r]{27.82}\hspace{0.05em}\makebox[2.7em][l]{\textcolor{black}{\tiny (\makebox[1.2ex][c]{-}\hphantom{0}0.34)}}} & \mbox{\makebox[2.3em][r]{\hphantom{0}3.67}\hspace{0.05em}\makebox[2.7em][l]{\textcolor{BrickRed}{\newthickdash{\textcolor{black}{\tiny (\makebox[1.2ex][c]{-}\hphantom{0}1.31)}}}}} \\
        Qwen 2.5 Coder 3B & \mbox{\makebox[2.3em][r]{28.30}\hspace{0.05em}\makebox[2.7em]{}} & \mbox{\makebox[2.3em][r]{\hphantom{0}5.23}\hspace{0.05em}\makebox[2.7em]{}} & \mbox{\makebox[2.3em][r]{28.44}\hspace{0.05em}\makebox[2.7em][l]{\textcolor{black}{\tiny (\makebox[1.2ex][c]{+}\hphantom{0}0.14)}}} & \mbox{\makebox[2.3em][r]{\hphantom{0}4.41}\hspace{0.05em}\makebox[2.7em][l]{\textcolor{black}{\tiny (\makebox[1.2ex][c]{-}\hphantom{0}0.82)}}} & \mbox{\makebox[2.3em][r]{28.25}\hspace{0.05em}\makebox[2.7em][l]{\textcolor{black}{\tiny (\makebox[1.2ex][c]{-}\hphantom{0}0.05)}}} & \mbox{\makebox[2.3em][r]{\hphantom{0}4.10}\hspace{0.05em}\makebox[2.7em][l]{\textcolor{black}{\tiny (\makebox[1.2ex][c]{-}\hphantom{0}1.13)}}} & \mbox{\makebox[2.3em][r]{33.08}\hspace{0.05em}\makebox[2.7em][l]{\textcolor{ForestGreen}{\uline{\textcolor{black}{\tiny (\makebox[1.2ex][c]{+}\hphantom{0}4.78)}}}}} & \mbox{\makebox[2.3em][r]{\hphantom{0}5.38}\hspace{0.05em}\makebox[2.7em][l]{\textcolor{ForestGreen}{\uline{\textcolor{black}{\tiny (\makebox[1.2ex][c]{+}\hphantom{0}0.15)}}}}} & \mbox{\makebox[2.3em][r]{28.78}\hspace{0.05em}\makebox[2.7em][l]{\textcolor{ForestGreen}{\uline{\textcolor{black}{\tiny (\makebox[1.2ex][c]{+}\hphantom{0}0.48)}}}}} & \mbox{\makebox[2.3em][r]{\hphantom{0}4.05}\hspace{0.05em}\makebox[2.7em][l]{\textcolor{BrickRed}{\newthickdash{\textcolor{black}{\tiny (\makebox[1.2ex][c]{-}\hphantom{0}1.18)}}}}} \\
        Qwen 2.5 Coder 7B & \mbox{\makebox[2.3em][r]{26.84}\hspace{0.05em}\makebox[2.7em]{}} & \mbox{\makebox[2.3em][r]{\hphantom{0}2.98}\hspace{0.05em}\makebox[2.7em]{}} & \mbox{\makebox[2.3em][r]{27.91}\hspace{0.05em}\makebox[2.7em][l]{\textcolor{black}{\tiny (\makebox[1.2ex][c]{+}\hphantom{0}1.07)}}} & \mbox{\makebox[2.3em][r]{\hphantom{0}3.76}\hspace{0.05em}\makebox[2.7em][l]{\textcolor{ForestGreen}{\uline{\textcolor{black}{\tiny (\makebox[1.2ex][c]{+}\hphantom{0}0.78)}}}}} & \mbox{\makebox[2.3em][r]{28.31}\hspace{0.05em}\makebox[2.7em][l]{\textcolor{ForestGreen}{\uline{\textcolor{black}{\tiny (\makebox[1.2ex][c]{+}\hphantom{0}1.47)}}}}} & \mbox{\makebox[2.3em][r]{\hphantom{0}3.61}\hspace{0.05em}\makebox[2.7em][l]{\textcolor{ForestGreen}{\uline{\textcolor{black}{\tiny (\makebox[1.2ex][c]{+}\hphantom{0}0.63)}}}}} & \mbox{\makebox[2.3em][r]{34.51}\hspace{0.05em}\makebox[2.7em][l]{\textcolor{ForestGreen}{\uline{\textcolor{black}{\tiny (\makebox[1.2ex][c]{+}\hphantom{0}7.67)}}}}} & \mbox{\makebox[2.3em][r]{\hphantom{0}5.56}\hspace{0.05em}\makebox[2.7em][l]{\textcolor{ForestGreen}{\uline{\textcolor{black}{\tiny (\makebox[1.2ex][c]{+}\hphantom{0}2.58)}}}}} & \mbox{\makebox[2.3em][r]{29.26}\hspace{0.05em}\makebox[2.7em][l]{\textcolor{ForestGreen}{\uline{\textcolor{black}{\tiny (\makebox[1.2ex][c]{+}\hphantom{0}2.42)}}}}} & \mbox{\makebox[2.3em][r]{\hphantom{0}4.07}\hspace{0.05em}\makebox[2.7em][l]{\textcolor{ForestGreen}{\uline{\textcolor{black}{\tiny (\makebox[1.2ex][c]{+}\hphantom{0}1.09)}}}}} \\[.5ex]
        \Xhline{2\arrayrulewidth}
        Mellum 4B Base - SFT & \mbox{\makebox[2.3em][r]{56.99}\hspace{0.05em}\makebox[2.7em]{}} & \mbox{\makebox[2.3em][r]{33.74}\hspace{0.05em}\makebox[2.7em]{}} & \mbox{\makebox[2.3em][r]{58.13}\hspace{0.05em}\makebox[2.7em][l]{\textcolor{ForestGreen}{\uline{\textcolor{black}{\tiny (\makebox[1.2ex][c]{+}\hphantom{0}1.14)}}}}} & \mbox{\makebox[2.3em][r]{35.32}\hspace{0.05em}\makebox[2.7em][l]{\textcolor{ForestGreen}{\uline{\textcolor{black}{\tiny (\makebox[1.2ex][c]{+}\hphantom{0}1.58)}}}}} & \mbox{\makebox[2.3em][r]{43.01}\hspace{0.05em}\makebox[2.7em][l]{\textcolor{BrickRed}{\newthickdash{\textcolor{black}{\tiny (\makebox[1.2ex][c]{-}13.98)}}}}} & \mbox{\makebox[2.3em][r]{24.27}\hspace{0.05em}\makebox[2.7em][l]{\textcolor{BrickRed}{\newthickdash{\textcolor{black}{\tiny (\makebox[1.2ex][c]{-}\hphantom{0}9.47)}}}}} & \mbox{\makebox[2.3em][r]{62.88}\hspace{0.05em}\makebox[2.7em][l]{\textcolor{ForestGreen}{\uline{\textcolor{black}{\tiny (\makebox[1.2ex][c]{+}\hphantom{0}5.89)}}}}} & \mbox{\makebox[2.3em][r]{44.12}\hspace{0.05em}\makebox[2.7em][l]{\textcolor{ForestGreen}{\uline{\textcolor{black}{\tiny (\makebox[1.2ex][c]{+}10.38)}}}}} & \mbox{\makebox[2.3em][r]{52.11}\hspace{0.05em}\makebox[2.7em][l]{\textcolor{BrickRed}{\newthickdash{\textcolor{black}{\tiny (\makebox[1.2ex][c]{-}\hphantom{0}4.88)}}}}} & \mbox{\makebox[2.3em][r]{31.90}\hspace{0.05em}\makebox[2.7em][l]{\textcolor{BrickRed}{\newthickdash{\textcolor{black}{\tiny (\makebox[1.2ex][c]{-}\hphantom{0}1.84)}}}}} \\
        Qwen 2.5 Coder 0.5B - SFT & \mbox{\makebox[2.3em][r]{49.45}\hspace{0.05em}\makebox[2.7em]{}} & \mbox{\makebox[2.3em][r]{26.13}\hspace{0.05em}\makebox[2.7em]{}} & \mbox{\makebox[2.3em][r]{50.59}\hspace{0.05em}\makebox[2.7em][l]{\textcolor{ForestGreen}{\uline{\textcolor{black}{\tiny (\makebox[1.2ex][c]{+}\hphantom{0}1.14)}}}}} & \mbox{\makebox[2.3em][r]{27.96}\hspace{0.05em}\makebox[2.7em][l]{\textcolor{ForestGreen}{\uline{\textcolor{black}{\tiny (\makebox[1.2ex][c]{+}\hphantom{0}1.83)}}}}} & \mbox{\makebox[2.3em][r]{50.21}\hspace{0.05em}\makebox[2.7em][l]{\textcolor{black}{\tiny (\makebox[1.2ex][c]{+}\hphantom{0}0.76)}}} & \mbox{\makebox[2.3em][r]{26.87}\hspace{0.05em}\makebox[2.7em][l]{\textcolor{black}{\tiny (\makebox[1.2ex][c]{+}\hphantom{0}0.74)}}} & \mbox{\makebox[2.3em][r]{65.90}\hspace{0.05em}\makebox[2.7em][l]{\textcolor{ForestGreen}{\uline{\textcolor{black}{\tiny (\makebox[1.2ex][c]{+}16.45)}}}}} & \mbox{\makebox[2.3em][r]{44.16}\hspace{0.05em}\makebox[2.7em][l]{\textcolor{ForestGreen}{\uline{\textcolor{black}{\tiny (\makebox[1.2ex][c]{+}18.03)}}}}} & \mbox{\makebox[2.3em][r]{55.45}\hspace{0.05em}\makebox[2.7em][l]{\textcolor{ForestGreen}{\uline{\textcolor{black}{\tiny (\makebox[1.2ex][c]{+}\hphantom{0}6.00)}}}}} & \mbox{\makebox[2.3em][r]{31.75}\hspace{0.05em}\makebox[2.7em][l]{\textcolor{ForestGreen}{\uline{\textcolor{black}{\tiny (\makebox[1.2ex][c]{+}\hphantom{0}5.62)}}}}} \\
        Qwen 2.5 Coder 1.5B - SFT & \mbox{\makebox[2.3em][r]{56.57}\hspace{0.05em}\makebox[2.7em]{}} & \mbox{\makebox[2.3em][r]{32.72}\hspace{0.05em}\makebox[2.7em]{}} & \mbox{\makebox[2.3em][r]{57.13}\hspace{0.05em}\makebox[2.7em][l]{\textcolor{black}{\tiny (\makebox[1.2ex][c]{+}\hphantom{0}0.56)}}} & \mbox{\makebox[2.3em][r]{33.89}\hspace{0.05em}\makebox[2.7em][l]{\textcolor{ForestGreen}{\uline{\textcolor{black}{\tiny (\makebox[1.2ex][c]{+}\hphantom{0}1.17)}}}}} & \mbox{\makebox[2.3em][r]{57.70}\hspace{0.05em}\makebox[2.7em][l]{\textcolor{black}{\tiny (\makebox[1.2ex][c]{+}\hphantom{0}1.13)}}} & \mbox{\makebox[2.3em][r]{35.30}\hspace{0.05em}\makebox[2.7em][l]{\textcolor{ForestGreen}{\uline{\textcolor{black}{\tiny (\makebox[1.2ex][c]{+}\hphantom{0}2.58)}}}}} & \mbox{\makebox[2.3em][r]{72.09}\hspace{0.05em}\makebox[2.7em][l]{\textcolor{ForestGreen}{\uline{\textcolor{black}{\tiny (\makebox[1.2ex][c]{+}15.52)}}}}} & \mbox{\makebox[2.3em][r]{52.99}\hspace{0.05em}\makebox[2.7em][l]{\textcolor{ForestGreen}{\uline{\textcolor{black}{\tiny (\makebox[1.2ex][c]{+}20.27)}}}}} & \mbox{\makebox[2.3em][r]{61.41}\hspace{0.05em}\makebox[2.7em][l]{\textcolor{ForestGreen}{\uline{\textcolor{black}{\tiny (\makebox[1.2ex][c]{+}\hphantom{0}4.84)}}}}} & \mbox{\makebox[2.3em][r]{39.65}\hspace{0.05em}\makebox[2.7em][l]{\textcolor{ForestGreen}{\uline{\textcolor{black}{\tiny (\makebox[1.2ex][c]{+}\hphantom{0}6.93)}}}}} \\
        Qwen 2.5 Coder 3B - SFT & \mbox{\makebox[2.3em][r]{58.69}\hspace{0.05em}\makebox[2.7em]{}} & \mbox{\makebox[2.3em][r]{35.20}\hspace{0.05em}\makebox[2.7em]{}} & \mbox{\makebox[2.3em][r]{61.24}\hspace{0.05em}\makebox[2.7em][l]{\textcolor{ForestGreen}{\uline{\textcolor{black}{\tiny (\makebox[1.2ex][c]{+}\hphantom{0}2.55)}}}}} & \mbox{\makebox[2.3em][r]{38.18}\hspace{0.05em}\makebox[2.7em][l]{\textcolor{ForestGreen}{\uline{\textcolor{black}{\tiny (\makebox[1.2ex][c]{+}\hphantom{0}2.98)}}}}} & \mbox{\makebox[2.3em][r]{63.43}\hspace{0.05em}\makebox[2.7em][l]{\textcolor{ForestGreen}{\uline{\textcolor{black}{\tiny (\makebox[1.2ex][c]{+}\hphantom{0}4.74)}}}}} & \mbox{\makebox[2.3em][r]{40.31}\hspace{0.05em}\makebox[2.7em][l]{\textcolor{ForestGreen}{\uline{\textcolor{black}{\tiny (\makebox[1.2ex][c]{+}\hphantom{0}5.11)}}}}} & \mbox{\makebox[2.3em][r]{77.81}\hspace{0.05em}\makebox[2.7em][l]{\textcolor{ForestGreen}{\uline{\textcolor{black}{\tiny (\makebox[1.2ex][c]{+}19.12)}}}}} & \mbox{\makebox[2.3em][r]{61.45}\hspace{0.05em}\makebox[2.7em][l]{\textcolor{ForestGreen}{\uline{\textcolor{black}{\tiny (\makebox[1.2ex][c]{+}26.25)}}}}} & \mbox{\makebox[2.3em][r]{67.54}\hspace{0.05em}\makebox[2.7em][l]{\textcolor{ForestGreen}{\uline{\textcolor{black}{\tiny (\makebox[1.2ex][c]{+}\hphantom{0}8.85)}}}}} & \mbox{\makebox[2.3em][r]{47.07}\hspace{0.05em}\makebox[2.7em][l]{\textcolor{ForestGreen}{\uline{\textcolor{black}{\tiny (\makebox[1.2ex][c]{+}11.87)}}}}} \\
        Qwen 2.5 Coder 7B - SFT & \mbox{\makebox[2.3em][r]{60.05}\hspace{0.05em}\makebox[2.7em]{}} & \mbox{\makebox[2.3em][r]{35.96}\hspace{0.05em}\makebox[2.7em]{}} & \mbox{\makebox[2.3em][r]{63.90}\hspace{0.05em}\makebox[2.7em][l]{\textcolor{ForestGreen}{\uline{\textcolor{black}{\tiny (\makebox[1.2ex][c]{+}\hphantom{0}3.85)}}}}} & \mbox{\makebox[2.3em][r]{41.18}\hspace{0.05em}\makebox[2.7em][l]{\textcolor{ForestGreen}{\uline{\textcolor{black}{\tiny (\makebox[1.2ex][c]{+}\hphantom{0}5.22)}}}}} & \mbox{\makebox[2.3em][r]{66.30}\hspace{0.05em}\makebox[2.7em][l]{\textcolor{ForestGreen}{\uline{\textcolor{black}{\tiny (\makebox[1.2ex][c]{+}\hphantom{0}6.25)}}}}} & \mbox{\makebox[2.3em][r]{44.14}\hspace{0.05em}\makebox[2.7em][l]{\textcolor{ForestGreen}{\uline{\textcolor{black}{\tiny (\makebox[1.2ex][c]{+}\hphantom{0}8.18)}}}}} & \mbox{\makebox[2.3em][r]{75.96}\hspace{0.05em}\makebox[2.7em][l]{\textcolor{ForestGreen}{\uline{\textcolor{black}{\tiny (\makebox[1.2ex][c]{+}15.91)}}}}} & \mbox{\makebox[2.3em][r]{58.99}\hspace{0.05em}\makebox[2.7em][l]{\textcolor{ForestGreen}{\uline{\textcolor{black}{\tiny (\makebox[1.2ex][c]{+}23.03)}}}}} & \mbox{\makebox[2.3em][r]{65.66}\hspace{0.05em}\makebox[2.7em][l]{\textcolor{ForestGreen}{\uline{\textcolor{black}{\tiny (\makebox[1.2ex][c]{+}\hphantom{0}5.61)}}}}} & \mbox{\makebox[2.3em][r]{44.38}\hspace{0.05em}\makebox[2.7em][l]{\textcolor{ForestGreen}{\uline{\textcolor{black}{\tiny (\makebox[1.2ex][c]{+}\hphantom{0}8.42)}}}}} \\[.5ex]
        \Xhline{2\arrayrulewidth}
        Qwen 3 Coder 480B A35B Instruct & \mbox{\makebox[2.3em][r]{52.52}\hspace{0.05em}\makebox[2.7em]{}} & \mbox{\makebox[2.3em][r]{29.10}\hspace{0.05em}\makebox[2.7em]{}} & \mbox{\makebox[2.3em][r]{57.37}\hspace{0.05em}\makebox[2.7em][l]{\textcolor{ForestGreen}{\uline{\textcolor{black}{\tiny (\makebox[1.2ex][c]{+}\hphantom{0}4.85)}}}}} & \mbox{\makebox[2.3em][r]{33.54}\hspace{0.05em}\makebox[2.7em][l]{\textcolor{ForestGreen}{\uline{\textcolor{black}{\tiny (\makebox[1.2ex][c]{+}\hphantom{0}4.44)}}}}} & \mbox{\makebox[2.3em][r]{61.12}\hspace{0.05em}\makebox[2.7em][l]{\textcolor{ForestGreen}{\uline{\textcolor{black}{\tiny (\makebox[1.2ex][c]{+}\hphantom{0}8.60)}}}}} & \mbox{\makebox[2.3em][r]{38.36}\hspace{0.05em}\makebox[2.7em][l]{\textcolor{ForestGreen}{\uline{\textcolor{black}{\tiny (\makebox[1.2ex][c]{+}\hphantom{0}9.26)}}}}} & \mbox{\makebox[2.3em][r]{69.36}\hspace{0.05em}\makebox[2.7em][l]{\textcolor{ForestGreen}{\uline{\textcolor{black}{\tiny (\makebox[1.2ex][c]{+}16.84)}}}}} & \mbox{\makebox[2.3em][r]{51.59}\hspace{0.05em}\makebox[2.7em][l]{\textcolor{ForestGreen}{\uline{\textcolor{black}{\tiny (\makebox[1.2ex][c]{+}22.49)}}}}} & \mbox{\makebox[2.3em][r]{62.70}\hspace{0.05em}\makebox[2.7em][l]{\textcolor{ForestGreen}{\uline{\textcolor{black}{\tiny (\makebox[1.2ex][c]{+}10.18)}}}}} & \mbox{\makebox[2.3em][r]{41.10}\hspace{0.05em}\makebox[2.7em][l]{\textcolor{ForestGreen}{\uline{\textcolor{black}{\tiny (\makebox[1.2ex][c]{+}12.00)}}}}} \\
        Claude 4.5 Sonnet & \mbox{\makebox[2.3em][r]{70.98}\hspace{0.05em}\makebox[2.7em]{}} & \mbox{\makebox[2.3em][r]{49.47}\hspace{0.05em}\makebox[2.7em]{}} & \mbox{\makebox[2.3em][r]{70.98}\hspace{0.05em}\makebox[2.7em][l]{\textcolor{black}{\tiny (\makebox[1.2ex][c]{ }\hphantom{0}0.00)}}} & \mbox{\makebox[2.3em][r]{49.47}\hspace{0.05em}\makebox[2.7em][l]{\textcolor{black}{\tiny (\makebox[1.2ex][c]{ }\hphantom{0}0.00)}}} & \mbox{\makebox[2.3em][r]{75.17}\hspace{0.05em}\makebox[2.7em][l]{\textcolor{ForestGreen}{\uline{\textcolor{black}{\tiny (\makebox[1.2ex][c]{+}\hphantom{0}4.19)}}}}} & \mbox{\makebox[2.3em][r]{54.73}\hspace{0.05em}\makebox[2.7em][l]{\textcolor{ForestGreen}{\uline{\textcolor{black}{\tiny (\makebox[1.2ex][c]{+}\hphantom{0}5.26)}}}}} & \mbox{\makebox[2.3em][r]{83.02}\hspace{0.05em}\makebox[2.7em][l]{\textcolor{ForestGreen}{\uline{\textcolor{black}{\tiny (\makebox[1.2ex][c]{+}12.04)}}}}} & \mbox{\makebox[2.3em][r]{70.52}\hspace{0.05em}\makebox[2.7em][l]{\textcolor{ForestGreen}{\uline{\textcolor{black}{\tiny (\makebox[1.2ex][c]{+}21.05)}}}}} & \mbox{\makebox[2.3em][r]{75.17}\hspace{0.05em}\makebox[2.7em][l]{\textcolor{ForestGreen}{\uline{\textcolor{black}{\tiny (\makebox[1.2ex][c]{+}\hphantom{0}4.19)}}}}} & \mbox{\makebox[2.3em][r]{54.73}\hspace{0.05em}\makebox[2.7em][l]{\textcolor{ForestGreen}{\uline{\textcolor{black}{\tiny (\makebox[1.2ex][c]{+}\hphantom{0}5.26)}}}}} \\[.5ex]
    \end{tabularx}
    \end{tcolorbox}
\end{table*}

%% file: latency.tex
\definecolor{veryLightGray}{RGB}{248,248,248}
\definecolor{midGray}{RGB}{220,220,220}

\renewcommand{\tabularxcolumn}[1]{m{#1}}

\begin{table}[thb]
    \caption{Inference Latency by Processor and Model\vspace{-0.35cm}}
    \label{tab:inference_latency}
    \footnotesize
    \begin{tcolorbox}[
        boxrule=0.5pt,
        colback=white,
        colframe=black!80,
        sharp corners,
        boxsep=0pt, left=0pt, right=0pt, top=0pt, bottom=0pt]
    \relsize{-1}
    \renewcommand{\arraystretch}{1.3}
    \setlength{\tabcolsep}{6pt}
    \begin{tabularx}{\linewidth}{l!{\vrule width 1pt}X|c|c}
        \rowcolor{black}
        \multicolumn{1}{l!{\color{white}\vrule width 1pt}}{\cellcolor{black}\textcolor{white}{\textbf{Processor}}} & \multicolumn{1}{l!{\color{white}\vrule}}{\cellcolor{black}\textcolor{white}{\textbf{Model}}} & \multicolumn{1}{c!{\color{white}\vrule}}{\cellcolor{black}\textcolor{white}{\textbf{Size}}} & \textcolor{white}{\textbf{Latency (s)}}\\
        \Xhline{1\arrayrulewidth}
        \multirow{2}{*}{M3 Max} & Qwen 2.5 Coder 3B - SFT & 5.75 GiB & $0.725$ \\
         & Qwen 2.5 Coder 7B - SFT (4-bit) & 4.36 GiB & $1.332$ \\
        \Xhline{2\arrayrulewidth}
        \multirow{2}{*}{M4 Max} & Qwen 2.5 Coder 3B - SFT & 5.75 GiB & $0.620$ \\
         & Qwen 2.5 Coder 7B - SFT (4-bit) & 4.36 GiB & $1.327$ \\
        \Xhline{2\arrayrulewidth}
        \multirow{2}{*}{RX 7800XT} & Qwen 2.5 Coder 3B - SFT & 5.75 GiB & $0.495$ \\
         & Qwen 2.5 Coder 7B - SFT (4-bit) & 4.36 GiB & $0.534$ \\
    \end{tabularx}
    \end{tcolorbox}
\end{table}

%% file: discussion.tex
\section{Discussion} \label{sec:dicussion}

\subsection{Problems, Findings, and Implications}

\subsubsection{What makes \pharo completion difficult?}
The main difficulty of \pharo completion is not simply data scarcity in isolation, but the combination of extremely low-resource conditions, Tonel-specific format, and Smalltalk-specific syntax. \pharo has far fewer public repositories than mainstream languages; Tonel mixes executable code with metadata; and \pharo's syntax differs substantially from high-resource languages in ways that hinder straightforward transfer. The method-level benchmarks failures are consistent with this: many failures arise from syntax issues such as keyword-message structure, precedence, delimiters, and exact AST boundaries, rather than the inability to solve the underlying programming task.

\subsubsection{On the value of specialization}

Specialization was very effective: our 3B and 7B models outperform far larger general-purpose models on the method-level \pharo benchmarks, and the specialized 7B model is also highly competitive at repository level. This matters because we target in-IDE completion rather than offline code generation. In that setting, latency and deployability are first-order concerns, motivating the use of compact models that can be quantized for real-time use. 
In repository-level experiments, context quality matters more than context quantity: the improvement from \textit{random methods} to \textit{impacted methods} show that semantically-related context is more effective than the mere addition of more repository data. This motivates further work in practical and effective context selection.

\subsubsection{On the engineering effort}

Specializing an LLM to a low-resource language such as \pharo involved a substantial engineering effort. We had to leverage domain knowledge about \pharo to: \begin{inparaenum}[(i)] \item identify and select which data we could use for training; \item develop tools to scrape, tokenize, parse, and evaluate \pharo code; \item construct domain-specific training datasets and evaluation benchmarks for \pharo; \item develop appropriate context selection policies for \pharo. \end{inparaenum} The effort involved in this significantly exceeded the effort in training the models themselves. The silver lining is that this investment can be reused when we need to specialize a newer LLM.

While other low-resource languages will have different idiosyncrasies than \pharo, we note that being able to parse the code was critical to many steps of the process (definition of training objectives, benchmark tasks, context selection). Further, the definition of language-specific benchmarks was a labor-intensive but critical element. In both cases, we think the strategies we employed can be applied for other languages.

\subsubsection{Comparison with ``low-resource'' literature}

Our work adds to the literature by improving LLMs on low-resource programming languages. LLMs can leverage transfer learning for tasks such as code generation \cite{Cass24a}, code translation \cite{Chen22a}, or program repair \cite{Prenner2023a}; other works have shown mixed evidence, indicating that pre-training is effective for small models, but less effective or potentially harmful for larger ones \cite{Giag25}. For low-resource code completion, evidence is limited: van Dam \etal \cite{VanD24a} evaluate UnixCoder and CodeGPT for Haskell code completion, while Gong \etal \cite{Gong22a} evaluate CodeGPT and a bespoke model, MultiCoder, for multiple languages, including Ruby. Both studies find that specialization is effective, but their findings apply to older and smaller models, with a limited context size. Our study complements these findings by evaluating larger, modern LLMs (yet still compatible with a completion scenario) on more challenging tasks, such as repository-level completion and completion tasks evaluated by test cases rather than similarity. Importantly, we find that specialized models can be competitive with much larger generalist models.

%% file: threats.tex
\subsection{Limitations} \label{sec:threats}

\subsubsection{Data contamination} We mitigate data contamination by using a June 1, 2024, cutoff for test repository selection and by removing from the training data all methods whose 8-grams overlap with the canonical solutions of the syntactic benchmarks. While this reduces the risk of data leakage, contamination cannot be fully ruled out, especially for the larger-model baselines.

\subsubsection{Method-level evaluation}
In this evaluation, we create masked spans from reference benchmark implementations. While this allows controlled evaluation with executable tests, the resulting tasks may not fully reflect the situations developers encounter during interactive coding. We partially mitigate this by simulating completions at arbitrary cursor positions, rather than AST boundaries, and via the repository-level evaluation.

Additionally, in this phase we only evaluate the continuous pre-trained models and their further fine-tuned version. Although a fine-tuned-only baseline could have provided an additional point of comparison, we did not consider it in this study as prior work \cite{Bava22} suggests that FIM capabilities are acquired more effectively during pre-training than through fine-tuning alone.

\subsubsection{Repository-level evaluation}
Unlike the method-level benchmarks, the repository-level evaluation lacks executable test cases. We use ChrF and CrystalBLEU, which measure closeness to the developer-written completion rather than functional correctness. These metrics are appropriate to evaluate similarity with real edits, but do not establish that a generated completion is semantically equivalent to the original one. In addition, the repository-level benchmark is derived from real commits and evaluated under several context policies, which strengthens realism. Nevertheless, it remains a simulation of editing behavior rather than a live IDE study in which developers would accept, edit or reject suggestions in practice.

\subsubsection{Inference evaluation}
The inference latency evaluation is limited to a small selection of machines with high specifications. While this is reasonable if one considers professional users, it does not cover all usage scenarios. Lower-end machines may benefit from cloud deployments. On the other hand, the latency evaluation is pessimistic, as we did not employ any efficient context selection strategy maximizing prompt caching to reduce latency.

%% file: conclusion.tex
\section{Conclusion \& Future Work} \label{sec:conclusion}

This study shows that bringing LLM-based code completion to a severely low-resource language such as \pharo is feasible. By combining \pharo-specific data curation with continued pre-training and fine-tuning, our pipeline substantially improves LLMs on both syntactic and repository-level benchmarks. In several settings, the resulting specialized models also match or exceed the accuracy of substantially larger general-purpose systems. Thus, for low-resource programming languages, specialization can compensate for scale while still yielding models compact enough for interactive IDE use. 
In future work, we plan to evaluate the models in realistic settings, including integration into the \pharo IDE and assessment under real user interaction. As part of that, we are currently developing a plugin that integrates LLM-based completion in the \pharo IDE. Finally, a second open question is how well our approach generalizes beyond \pharo to other Smalltalk dialects.